\documentclass[a4paper,11pt]{article}
\usepackage{jcappub} 
\usepackage{lineno}
\usepackage{bm}
\usepackage{makecell}
\usepackage{csvsimple}
\usepackage{comment}
\usepackage{natbib}

\arxivnumber{------------} 
\title{Systematics in ETG Mass Profile Modelling: Strong Lensing \& Stellar Dynamics}

\author[a]{Carlos R. Melo-Carneiro}
\author[a]{Cristina Furlanetto}
\author[a]{Ana L. Chies-Santos}
\affiliation[a]{Instituto de Física, Universidade Federal do Rio Grande do Sul, \\
Av. Bento Gonçalves 9500 - Porto Alegre, R.S. 90040-060, Brazil}

\emailAdd{carlos.melo@ufrgs.br}

\abstract{Strong gravitational lensing and stellar dynamics are independent and powerful methods to probe the total gravitational potential of galaxies, and thus, their total mass profile. However, inherent degeneracies in the individual models makes it difficult to obtain a full understanding of the distribution of baryons and dark matter (DM), although such degeneracies might be broken by the combination of these two tracers, leading to more reliable measurements of the mass distribution of the lens galaxy. We use mock data from IllustrisTNG50 to compare how dynamical-only, lens-only, and joint modelling can constrain the mass distribution of early-type galaxies (ETGs). The joint model consistently outperforms the other models, achieving a $2\%$ accuracy in recovering the total mass within $2.5R_\text{eff}$. The Einstein radius is robustly recovered for both lens-only and joint models, with the first showing a median fractional error of $-5\%$ and the latter a fractional error consistent with zero. The stellar mass-to-light ratio and total mass density slope are well recovered by all models. In particular, the dynamical-only model achieves an accuracy of $1\%$ for the stellar mass-to-light ratio, while the accuracy of the mass density slope is typically of the order of $5\%$ for all models. However, all models struggle to constrain integrated quantities involving DM and the halo parameters. Nevertheless, imposing more restrictive assumptions on the DM halo, such as fixing the scale radius, could alleviate some of the issues. Finally, we verify that the number of kinematical constraints ($15, 35, 55$ bins) on the kinematical map does not impact the models outcomes. \\ \\ keywords: gravitational lensing - galaxy dynamics - hydrodynamical simulations }

\begin{document}
\maketitle
\flushbottom

\section{Introduction} \label{sec:intro}

Within the $\Lambda$-cold dark matter ($\Lambda$CDM) model, the current cosmological framework, galaxies are described as being gravitationally bound systems primarily composed of stars and gas (baryons) alongside a massive halo of non-luminous matter known as dark matter (DM). Understanding the formation and evolution of such systems is crucial for astrophysics and cosmology. Of particular interest is the formation and evolution of early-type galaxies (ETGs), recognised as the end product of a galaxy's life. Their evolution is currently well-described by a two-phase scenario (e.g., \citep[][]{Oser2010,Rodriguez-Gomez2016}), starting from an active phase of \textit{in-situ} star formation and gas-rich mergers at redshift $z \geq 2$, followed by a passive phase ($z \leq 2$) dominated by the accretion of \textit{ex-situ} formed stars and dry mergers (though, see \cite{Wang2019} for a recent claim proposing a three-phase evolutionary scenario). 

It is widely recognised that the mass distribution of ETGs reflects their assembly history. The total matter density distribution of ETGs is well approximated by a power-law model, $\rho_{\text{total}}(r) \propto r^{\gamma}$, even though their baryonic and DM components do not individually follow this distribution. Simulations \citep[][]{Remus2017, Wang2019} indicate that the slope of this distribution reflects the two-phase evolutionary scenario. At higher redshifts ($z \sim 2-3$), the slope is steeper than isothermal ($\gamma = 2$) due to dissipative processes such as wet mergers. After peaking at $z \sim 2$, the slope decreases between $z \sim 1-2$, with main progenitors evolving passively into ETGs, and $\gamma$ remains nearly invariant, becoming near-isothermal close to $z=0$. This near-isothermal state ($\gamma \sim 2$) is debated, since baryonic and DM density profiles differ significantly from an isothermal distribution. However, the total mass distribution appears to ``conspire'' to this form \citep[][]{Treu&Koopmans2004, Ruff2011, Serra2016}. This fact has been highlighted by different observational studies: dynamical modelling of local ETGs \citep[][]{Tortora2014a,Cappellari2015-density,Poci2017,Li2019}, as well as intermediate ($0.29<z<0.55$) \citep[][]{Derkenne2021}, weak lensing observations \citep[][]{Gavazzi2007}, strong gravitational lensing (SLG) modelling \citep[][]{Etherington2023}, and combined SGL and dynamical modelling \citep[][]{Koopmans2006,Koopmans2009,Barnabe2011,Sonnenfeld2013,Li2018}.

Beyond the total mass distribution, disentangling the baryonic and DM distributions is crucial for better understanding galaxy evolution processes. The central DM fraction is connected with the galaxy evolution growth (e.g., \citep[][]{Tortora2012,Tortora2014b,Lovell2018}), as well as the stellar content within the galaxy (e.g., \cite{Cappellari2013,Tortora2014b}. The shape of the DM profile itself may provide crucial information about the interplay of baryons and DM (e.g., \cite{Gnedin2004,Anbajagane2022,Petit2023}) and the presence of feedback mechanisms (e.g., \cite{Navarro1996,Duffy2010,Cintio2014,Jackson2023}). In particular, this latter could be intrinsically connected with the ``cusp-core'' problem (see ref. \citep{Popolo2022} for a review) that challenges the $\Lambda$CDM model. 

To effectively disentangle the stellar and DM components, it is essential to accurately determine both mass distributions. However, we typically only have access to the total density distribution through modelling, which turns such disentangling challenging \cite{Sonnenfeld2018, Liang2023}. This challenge arises because, although the stellar component can be directly probed by light (in contrast to the DM counterpart), converting the observed luminosity $L$ to the stellar mass $M_\text{star}$, depends on the determination of the stellar mass-to-light ratio $\Upsilon_\star = M_\text{star}/L$. Such determination, often made through stellar population synthesis (SPS), relies on the initial mass function (IMF), which, in turn, depends on accurate calibrations related to $\Upsilon_\star$. Hence, independent determinations of $\Upsilon_\star$ must be conducted. In the case of galaxies beyond the local Universe ($z \gtrsim 0.2$), where high-quality spectroscopic observations are not easily available, the determination of $\Upsilon_\star$ often depends on stellar/gas dynamical modelling (e.g., \cite{Cappellari2012, Davis2017}) and SGL (e.g., \cite{Sonnenfeld2012, Sonnenfeld2015, Newman2017}). The stellar masses derived from these methods are independent of the IMF assumptions, allowing them to be used to infer the appropriated IMFs and ultimately compare them to those obtained from the SPS modelling.

Therefore, investigations of the inner mass distribution of ETGs at various redshifts are of utmost importance to address numerous issues related to their formation and evolution. For nearby systems ($z \lesssim 0.1$), stellar dynamical modelling has proven effective in providing comprehensive insights into the mass structure of ETGs (e.g., \cite{Krajnovic2005,Li2019, Santucci2022, Zhu2023}). However, this technique faces challenges when applied to more distant objects, due to the same issues mentioned above. In this more distant regime, SGL takes place as an accurate and independent method for measuring the total projected mass profile (\cite{Dye2005, Nightingale2019, Shajib2022}). Nonetheless, despite the great success of the individual methods, both are susceptible to degeneracies that can prevent a precise recovery of the mass profile. Namely, SGL modelling is affected by the so-called mass-sheet degeneracy \citep[][]{Gorenstein1988,Schneider2013}, while the stellar dynamics is subject to the mass-anisotropy degeneracy \citep[][]{Gerhard1993,Read2017}.

A well-known and extensively explored solution to break these degeneracies is the joint modelling of SGL and stellar dynamics (see, e.g., the refs. \cite{Treu&Koopmans2004, Koopmans2006, Koopmans2009, Auger2010, Barnabe2009, Barnabe2012, Sonnenfeld2013, Li2018, Tan2023}). However, the majority of these works treat SGL and stellar dynamics as independent problems, where the projected mass profile obtained from the lens modelling is used as a prior constraint for the dynamical model or vice-versa. Another issue in many of these previous works is the reliance on a single velocity dispersion measurement for the dynamical model, typically acquired from integrated spectroscopic observations. Some studies that overcome both of these issues are the works by refs. \cite{Barnabe2009, Barnabe2012}, where fully self-consistent modelling of SGL and stellar dynamics is employed. In these works, besides the photometric data necessary for SGL analyses, the authors also use spatially resolved stellar kinematical maps extracted from integral field units (IFU) for the dynamical modelling. This self-consistent model applied to photometric+IFU observations could have some advantages over the traditional methods, once the galaxy mass profile can be constrained at different radii. Nevertheless, a self-consistent mass profile represents a more rigorous physical framework, in the sense that the same mass model should reproduce both observational datasets simultaneously.

The high-quality image and spectroscopic data from current large astronomical surveys motivates a fresh view of this kind of self-consistent modelling, which is the goal of the present paper. The proposed method studied here solves, for a given mass distribution and dataset (i.e., the observed lensed galaxy, the lens surface brightness, and lens line-of-sight (LOS) projected velocity dispersion map), the axisymmetric Jeans equations \citep[][]{Jeans1915,Binney2008} for stellar dynamics and the lens equation \citep[][]{Schneider1992} for the lensing phenomena in a self-consistent manner. Through a Bayesian framework, the posterior distribution of parameters describing the gravitational potential of the lens galaxy can be determined, enabling the recovery of the underlying mass distribution and the reconstruction of the lensed galaxy. 

In this paper, we aim to assess the robustness of the joint modelling of SGL and stellar dynamics in recovering the mass properties of the lens galaxy, such as the stellar mass, DM mass, DM fraction, and total mass. We also investigate how accurately the method recovers the total mass density slope, the DM parameters, and if the number of kinematical constraints impacts the model outcomes. Finally, we compare these results with results obtained by dynamical-only and lens-only modelling. To achieve these goals, we applied the method to mock galaxies obtained from a hydrodynamical cosmological simulation. This methodology enables us to compare the complex mass distribution of the mock ETG, serving as a lens, with the mass profile derived through the joint modelling.

This paper is structured as follows. In section \ref{sec:sample}, we introduce the mock sample and how we simulate the observation from the particle data. In section \ref{sec:theory}, we present an overview of the theory and modelling process, including the description of the mass profile and pipeline applied in this work. In section \ref{sec:results} we show the results of the joint modelling and the comparison with individual models of SGL and/or stellar dynamics. Then we summarise and conclude in section \ref{sec:conclusions}. 

Throughout this paper, unless explicitly stated otherwise, we adopt the cosmological parameters consistent with ref. \cite{Planck_2015} results: $\Omega_{\Lambda,0} = 0.6911$, $\Omega_{\text{m},0} = 0.3089$, $\Omega_{\text{b},0} = 0.0486$, and $H_0 = 67.74$ km s$^{-1}$ Mpc$^{-1}$. 

\section{Mock Sample} \label{sec:sample}
\subsection{The IllustrisTNG simulation}

To create the mock lens galaxy sample, we used data from the cosmological hydrodynamical simulations IllustrisTNG-50 (\cite{TNG}, TNG50 hereafter). This simulation offers a remarkable combination of volume and resolution, representing the highest resolution realisation within the IllustrisTNG project \citep{Springel2018,Pillepich2018b,Nelson2019}, running in a periodic-boundary cube with a comoving size of $51.7$ Mpc and softening length scale at redshift zero of $\epsilon^{z=0}_{\text{DM,stars}} = 288$\,pc for the collisionless components.

TNG50 is a gravo-magnetohydrodynamical simulation performed using the moving-mesh code \texttt{AREPO} \citep[][]{Springel2010}, which also incorporates a galaxy sub-grid model, including different AGN feedback modes, chemical enrichment, stellar feedback, and other baryonic processes \citep[][]{Weinberger2017,Pillepich2018}. The simulation evolves DM particles, stars, gas, black holes, and magnetic fields from a redshift of $z = 127$ until $z = 0$. The initial conditions are set based on cosmological motivations and follow the cosmological parameters consistent with the results of ref. \cite{Planck_2015} (the same employed in this work). The main features of the TNG50 run, such as the mass resolution and softening length scale of the particles, are summarised in ref. \cite{Nelson2019}, Table 1.  

The particle data are organised into ``snapshots'', which represent different redshifts in the simulated Universe. In each snapshot, halos are identified using a Friends-of-Friends (\texttt{FoF}) algorithm, while the structures called subhalos (the analogous of galaxies in the simulation) are identified using the \texttt{Subfind} algorithm \citep[][]{Springel2001,Dolag2009}. Merger trees are created using the \texttt{SubLink} algorithm \citep[][]{Rodriguez-Gomez2015}.

\subsection{Sample selection}
Following the Sloan Lens ACS Survey (\cite{Bolton2008-slacs}, SLACS hereafter), for which the median lens redshift is  $z_l \sim 0.2$, we looked for subhalos in the TNG50 simulation at snapshot 84, which corresponds to $z = 0.2$ (angular diameter distance $702$\,Mpc) in the simulated Universe. 

Given the minimal evolution of ETGs between $z=1$ and $z=0$ (e.g., \cite{Koopmans2006, Koopmans2009, Sonnenfeld2013, WangYunchong2022}), we opted to use available catalogues that had previously classified the subhalos according to their morphology and track the subhalos to the desired snapshot to match the SLACS median lens redshift. We use the SKIRT Synthetic Images and Optical Morphologies catalogue \citep[][]{Rodriguez-Gomez2019} and the Galaxy Morphologies (Deep Learning) catalogue \citep[][]{Huertas-Company2019,Varma2022}, with the intent of selecting sufficient candidates. The first catalogue builds optical images (SDSS and Pan-STARRS filters) for all subhalos with stellar mass $M_{\text{star}} > 10^{9.5}$\,M$_\odot$ at redshifts $z=0$ and $z=0.05$, and then fit the images with a Sérsic profile. The second catalogue uses a convolutional neural network trained on visual morphologies from the SDSS to classify mock SDSS images of subhalos with stellar mass $M_{\text{star}} > 10^{9}$\,M$_\odot$ and redshifts $z=0.5, 1, 1.5, 2, 2.5, 3$.

For the \cite{Rodriguez-Gomez2019} catalogue, we use the SDSS r-band information at redshift $z=0.05$ to identify potential ETGs. We apply four queries to this catalogue: 1) {\fontfamily{qcr}\selectfont flag $=0$}, to ensure reliable morphological measurements; 2) {\fontfamily{qcr}\selectfont flag\_sersic $=0$}, to ensure reliable Sérsic parameters; 3) {\fontfamily{qcr}\selectfont sn\_per\_pixel $>5$}, which means that the image used to derive the Sérsic profile has a signal-to-noise ratio (SNR) greater than $5$; and 4) {\fontfamily{qcr}\selectfont sersic\_n $>2.5$}, to select only subhalos fitted with a Sérsic index greater than $2.5$, a typical value for ETGs.

The second catalogue provides probabilities for the image of each subhalo to have a spheroid-like morphology ({\fontfamily{qcr}\selectfont P\_Spheroid}), a disk-like morphology ({\fontfamily{qcr}\selectfont P\_Disk}), or an irregular-like morphology ({\fontfamily{qcr}\selectfont P\_Irr}). In this case, we focus on the classifications corresponding to redshift $z=0.5$ and select subhalos with {\fontfamily{qcr}\selectfont P\_Spheroid $>0.75$} and {\fontfamily{qcr}\selectfont P\_Disk $<0.25$}.

For each subhalo selected previously, we use the Python module \texttt{ILLUSTRIS\_PYTHON}\footnote{\url{https://github.com/illustristng/illustris\_python}}, to track the subhalos up to redshift $z=0.2$ using their merger trees. This implies that, for the first selection made at $z=0.05$, we track the progenitors of each subhalo, while for the subhalos selected at $z=0.5$ we track their decedents.

After tracking all the selected subhalos to snapshot 84, the following queries were applied: 1) $M_{\text{star}} > 10^{9.5}$\,M$_\odot$ within $10$\,kpc, which corresponds to approximately $3$\,arcsec at $z=0.2$; 2) the stellar mass should be $2$\,dex greater than the gas mass within  $10$\,kpc; and 3) $M_{\text{DM}} > 10^{10.5}$\,M$_\odot$, including all DM particles in the subhalo.

The first and third criteria were chosen to ensure a good number of stellar and DM particles, while the second one is motivated to avoid subhalos where the gas mass fraction is significant. Those with distinctive arm-like structures after projection (see Sec. \ref{subsec:mock}) or lacking sufficient DM particles for the construction of a reliable radial density profile were excluded from the remaining subhalos. As a result, we carefully select $21$ subhalos that represent the final sample. Appendix \ref{ap:sample} provides further details on their properties.

\subsection{Mock observations}\label{subsec:mock}
To generate the mock observations from the particle data, we followed a similar approach as described in ref. \cite{Li2016} and implemented as the public code \texttt{illustris-tools}\footnote{\url{https://github.com/HongyuLi2016/illustris-tools}}, although with some
updates to fit our TNG50 sample and lensing mock observations.

From now on, we will consider that the total mass distribution of the subhalos is composed of their stellar plus DM components only. This is motivated by the fact that we only selected subhalos whose gas mass fraction is negligible. We also consider primed notation as projected coordinates in the sky plane and non-primed as intrinsic coordinates. In particular, $(x^{\prime}, y^{\prime}, z^{\prime})$ represents Cartesian coordinates assuming that the $z^{\prime}$-axis aligns with the LOS and $x^{\prime}$-axis aligned with the galaxy's projected semi-major axis. Furthermore, we will use the notation $\mathbf{U}[a, b]$ for a uniform prior with a lower value $a$ and an upper value $b$; $\mathcal{N}[a,b]$ for a Gaussian normal prior with mean $a$ and dispersion $b$; and $\log_{10}\mathbf{U}[a,b]$ for a log-uniform distribution with a lower value $a$ and a maximum value $b$.

\subsubsection{Lens shape and projection}
We determined the centre of mass for each particle set (stellar and DM) by iteratively refining $20\%$ of the particles in the central region based on their radial distances from the current centre. This process stops when the difference between consecutive centre estimates falls below $0.01$\,kpc. The velocity of the centre of mass is then calculated as the mass-weighted average of particles within $15$\,kpc. Finally, all particle coordinates and velocities are calibrated to these central values.

To make 2D projections of the particles and mimic observations, we determine the intrinsic shape of the subhalos using the reduced inertia tensor method \citep[][]{Allgood2006} to find their principal axes. The inertia tensor is defined as

\begin{equation}
I_{i,j} = \sum_k \frac{x^{(k)}_i x^{(k)}_j}{r^2_k},
\end{equation}
where $r_k$ is the radial distance from the centre of mass to the $k^{\text{th}}$ particle, $x^{(k)}_i$ is the position of the $k^{\text{th}}$ particle, and the sum is performed over a set of $k$ particles of interest.

Assuming that the galaxy can be represented by an ellipsoid with axes $a \geq b \geq c$, the axis ratios defined by $q = b/a$ and $s = c/a$, can be obtained from the ratio of the square root of the eigenvalues of the inertia tensor, $(a,b,c) = \sqrt{(\lambda_a,\lambda_b,\lambda_c)}$, where $\lambda_i$ is one of the eigenvalues. Initially, the method considers all particles within a sphere of $30$\,kpc to the centre of mass to define the inertia tensor. The distances are calculated as $r^2_k = x^2_k + y^2_k/q^2 + z^2_k/s^2$, assuming $q = s = 1$ at the beginning. Then, the method iteratively refines the galaxy's shape by: 1. Diagonalising the inertia tensor: Finds the eigenvalues and the principal axes of the inertia tensor; 2. Updating the axis ratios: Calculates new values for $q$ and $s$ based on the updated eigenvalues; 3. Deforming the sphere: Stretches or shrinks the initial sphere along the principal axes according to the updated ratios; 4. Convergence check: Repeats steps 1-3 until the difference between consecutive $q$ or $s$ values is smaller than $1\%$.

After determining the galaxy's shape, the particles are rotated such that the $x$-axis is aligned with the longest axis and the $z$-axis is aligned with the shortest axis. Optionally, a rotation $\phi$ along the $z$-axis is allowed before performing the projection along the inclination angle $i$, which represents the angle between the shortest axis and the LOS ($i = 90^{\circ}$ implies that the galaxy is seen edge-on). After the projection, an additional rotation by the position angle (in the sky-plane) is allowed. 

\subsubsection{Lensing data}
According to our sample definition, the lens galaxies are at $z_l = 0.2$. Then we project the mass particles of a given galaxy onto a square grid with a side length of $80$\,kpc and a pixel scale of $0.09$\,arcsec. The choice of pixel size was made to resemble the resolution of the Hubble Space Telescope (HST) Wide Field Camera 3 (WFC3). This projection yields the surface mass density of the galaxy, and we create separate projections for the stellar, DM and total (stars+DM) mass. The inclination angle ($i$) and the rotation along the $z-$axis ($\phi$) are both randomly sampled from uniform distributions: $i \sim \mathbf{U}[65^{\circ},85^{\circ}]$ and $\phi \sim \mathbf{U}[0^{\circ},180^{\circ}]$. After that, the position angle is determined such that the galaxy's semi-major axis aligns with the $x^\prime$-axis.

We convert the mock galaxy's stellar surface mass profile to the surface brightness profile by dividing the former by a constant stellar mass-to-light ratio $\Upsilon_\star$ for simplicity. The mass-to-light ratio is defined using the SDSS r-band luminosity provided in the header of each subhalo, which can be accessed through the {\fontfamily{qcr}\selectfont GFM\_StellarPhotometrics} flag\footnote{Further details can be found at \url{https://www.illustris-project.org/data/forum/topic/445/gfm\_stellarphotometrics-and-model-a-in-nelson-2018/}}. We calculate the ratio of the total mass to the total luminosity of all stellar particles within the subhalo. The specific method used to establish $\Upsilon_\star$ is irrelevant, and our chosen procedure is merely intended to introduce some variance in the values of the stellar mass-to-light ratios.

Using the total surface mass density, we follow the approach of ref. \cite{Cao2022} to create the lensing mock data. We create the correspondent convergence map (see details in section \ref{sec:theory}), from which the deflection field is obtained by the \texttt{lenstronomy} \citep[][]{Birrer2018,Birrer2021} routine {\fontfamily{qcr}\selectfont deflection\_from\_kappa\_grid}. We adopt a Sérsic profile \citep[][]{Sersic1968} for the source galaxy:

\begin{equation}\label{eq:sersic}
    I_{\text{Ser}}(\xi) = I_{\text{source}} \exp\left\{ -k\left[ \left( \frac{\xi}{R_{\text{eff}}}\right)^{\frac{1}{n}} - 1\right]     \right\},
\end{equation}
where $\xi = \sqrt{ {x^\prime}^2_0 + {y^\prime}^2_0/q^2_{\text{source}}}$ is an elliptical coordinate with ($x^\prime_0, y^\prime_0$) being the source light centre, $q_{\text{source}}$ is the source axial ratio, $I_{\text{source}}$ is the source intensity, $R_{\text{eff}}$ is the effective radius, $n$ is the Sérsic index, and $k$ is a function of $n$. Additionally, the light profile has a position angle $\phi_{\text{source}}$ defined counterclockwise from the positive $x^\prime$-axis. 

The source parameters are randomly drawn, for each mock lens system, from the following distributions: ($x^\prime_0, y^\prime_0$) $\sim \mathcal{N}[0.0,0.1]$\,arcsec; $q_{\text{source}} \sim \mathbf{U}[0.5,1.0]$; $\phi_{\text{source}} \sim \mathbf{U}[0^{\circ},180^{\circ}]$; $R_{\text{eff}} \sim \mathbf{U}[0.01,3.0]$\,arcsec; $n \sim \mathbf{U}[1.5,3.5]$. Additionally, we regulate the intensity to maintain an SNR of approximately $50$ in the brightest pixel. The source's redshift is sampled from the source redshift distribution of the SLACS sample \citep[][]{Bolton2008-slacs}, satisfying the constraint $z_s > z_l$, where $z_s$ is the source redshift. The lensed image is then convolved with a Gaussian point-spread function (PSF) with dispersion of $0.05$\,arcsec. We also take into account a sky background of $0.1$\,counts\,s$^{-1}$ and an exposure time of $840$\,s, from which a Poisson noise is added. These values are similar to those expected for an HST-like observation.

Since this work focuses on evaluating the systematic errors in the combined modelling of SGL and stellar dynamics, we intentionally avoid the inclusion of the lens surface brightness profile in the observed mock image data to mitigate other potential sources of systematic. The deflected sources can be seen in the first column of Figures \ref{fig:TNG50_sample}.

\subsubsection{Kinematical data}

The mock IFU kinematical maps were constructed using the same projection as above. However, for the kinematical data, we use a grid with a pixel scale of $0.2$\,arcsec to mimic the Multi-Unit Spectroscopic Explorer (\cite{2010SPIE.7735E..08B}, MUSE hereafter) spatial resolution. To simulate an IFU aperture, only the central region satisfying

\begin{equation}
    r^{\prime} = \sqrt{ {x^\prime}^{2} + \frac{{y^\prime}^{2}}{{q^\prime}^{2}} } < \frac{2.5R_{\text{eff}}}{\sqrt{q^\prime}},
\end{equation}
were used to construct the kinematical maps. Here $R_{\text{eff}}$ is the effective radius determined by the Multi-Gaussian Expansion (MGE, \cite{Emsellem1994E,Cappellari2002}) model (see details in section \ref{sec:theory}), and $q^\prime$ is the galaxy observed axial ratio, also determined during the MGE fit.

Each selected pixel is then considered an IFU spaxel. By using the Convex-Hull method, stellar particles projected within the IFU aperture are assigned to their nearest spaxel by querying the KD-Tree constructed for all spaxel anchor points \citep[][]{WangYunchong2022}. Finally, the spaxels are Voronoi-binned using the Python package \texttt{VorBin} \citep[][]{Cappellari2003} to achieve a minimum SRN of $50$ particles per spaxel, which ensures good quality kinematic information. In each Voronoi bin, the mean velocity ($\overline{v_{\text{LOS}}}$) and the velocity dispersion ($\sigma^2_{\text{LOS}}$) along the LOS are obtained by the mass-weighted mean velocity and the mass-weighted mean second velocity moment ($\overline{v^2_{\text{LOS}}}$) of the stellar particles. The velocity dispersion is obtained by $\sigma^2_{\text{LOS}} = \overline{v^2_{\text{LOS}}} - \overline{v_{\text{LOS}}}^2$. Finally, the second velocity moment is treated as the stellar kinematic observable, once it can be directly compared with the root-mean-square velocity $v_{\text{rms}} = \sqrt{v^2 + \sigma^2_v}$, where $v$ and $\sigma_v$ are the observed LOS stellar mean velocity and velocity dispersion, respectively. 

During the Voronoi-binning of the IFU mock data, the number of Voronoi bins were regulated to a total of $35$, each containing a minimum of $50$ particles. We also create two more variations of the IFU mock data, one with $15$ bins and another with $55$. These variations are discussed in section \ref{sec:results}, and they are used to assessing possible systematics related to the number of kinematical tracers.

To ensure more realistic uncertainties in the kinematical measurements, we derive them using a simulation approach. First, we use SSP models to create $144$ simulated galaxy spectra, varying parameters such as age, metallicity, and SNR. Subsequently, each simulated spectrum was convolved with a LOS velocity distribution and the MUSE line spread function, as characterised by ref. \cite{Guerou2017}, which accounts for the instrumental spectral resolution. Finally, we employ the Penalized PiXel-Fitting (\texttt{pPXF}, \cite{Cappellari2004,Cappellari2017}) algorithm to fit the first two velocity moments of all $144$ spectra using the Indo-US \citep{Indo-US} stellar library within the rest frame range of $3800-5300$\,\AA. For each spectral fit, we determine the uncertainty in the fitted parameters by employing a Monte Carlo approach with $200$ realisations, and then we propagate the uncertainties to $v_{\text{rms}}$ using

\begin{equation}\label{eq:error_vrms}
    1\sigma_{\text{rms}} = \frac{\sqrt{ (v \times 1\sigma_{\text{vel}})^2 + (\sigma_v \times 1\sigma_{\text{disp}})^2   }}{v_{\text{rms}}},
\end{equation}
where $1\sigma_{\text{rms}}$ is the  uncertainty in the rms velocity, $1\sigma_{\text{vel}}$ and $1\sigma_{\text{disp}}$  are the $1\sigma$ uncertainty on the velocity and velocity dispersion, respectively.

We observed that regardless of the configurations of the simulated spectra, the uncertainty in the $v_{\text{rms}}$ consistently ranges between $9\%$ and $14\%$ of its true value. The histogram illustrated in Figure~\ref{fig:vrms_dist_u} displays the ratio of the uncertainty in $v_{\text{rms}}$ to the true $v_{\text{rms}}$ value for the $144$ simulated spectra. Considering the minimal variation in the uncertainty of $v_{\text{rms}}$ across different SNR levels, ages, metallicity, and $\sigma_v$ values, we adopt the median value of $11\%$ of the $v_{\text{rms}}$ value as the uncertainty for each Voronoi bin in the mock IFU observations. Further details of the spectral simulation and derived uncertainties are given in Appendix \ref{ap:kin_unc}. 

\subsubsection{DM radial profiles}
We create a radial mass density profile for the DM component to compare it with the model predictions. The radial mass distribution is determined by averaging the particles within spherical shells. First, we calculate the half-mass radius using the mass growth curve of all DM particles within the subhalo. The mass growth curve is computed within a range of $100$ logarithmically spaced spherical bins, spanning from $r_{\text{min}} = 2.8\epsilon^{z=0}_{\text{DM,stars}}$ to $r_{\text{max}}$, where $r_{\text{max}}$ is the maximum radius determined by the particles. Subsequently, we determine the half-mass radius through linear interpolation of the mass growth curve, as done by ref. \citep[][]{Petit2023}.

After determining the DM half-mass radius, the DM mass density distribution is computed by averaging particles within $100$ logarithmically spaced spherical shells ranging from $r_{\text{min}} = 2.8\epsilon^{z=0}_{\text{DM,stars}}$ to the half-mass radius. This logarithmic spacing ensures narrower central bins where the density of particles is higher and wider outermost bins where the density of particles is lower. The minimum value represents the limit where gravitational forces behave according to Newtonian physics \citep{Pillepich2018}. Finally, the DM radial density is fitted using a generalised Navarro-Frenk-White (gNFW; \cite{Wyithe2001}) profile :

\begin{equation}\label{eq:gNFW}
    \rho(r) = \rho_s  \left( \frac{r}{r_s} \right)^{-\gamma_{\text{DM}}} \left(1 + \frac{r}{r_s}\right)^{\gamma_{\text{DM}} - 3},
\end{equation}
where $\rho_s$ is a characteristic density at the scale radius $r_s$, and $\gamma_{\text{DM}}$ is the inner density slope that allows the profile to be cuspier ($\gamma_{\text{DM}} > 1$) or cored ($\gamma_{\text{DM}} = 0$). When $\gamma_{\text{DM}} = 1$ the profile reduces to the classical NFW \citep[][]{Navarro1997}.

The fit is performed using the nested sampling approach as implemented in the \texttt{dynesty} algorithm \citep[][]{Speagle2020,sergey_koposov_2023_7995596}. The posterior distributions of the gNFW parameters are obtained using the \texttt{dynesty} default configuration with the following priors: $\log(\frac{\rho_s}{\text{M$_\odot$pc$^{-3}$}}) \sim \mathbf{U}[-6,0]$, $r_s \sim \mathbf{U}[0.01,30]$\,arcsec and $\gamma_{\text{DM}} \sim \mathbf{U}[0.5,2]$. The sampling stop criteria is $\Delta \ln \hat{\mathcal{Z}} = 0.8$, where $\Delta \ln \hat{\mathcal{Z}}$ is the estimated remaining Bayes evidence. 

\subsection{Caveats to the mock observations}
Despite the care we take to create the mock observations, some caveats must be highlighted.

First, the majority of the lensing data we generate exhibit a small central image, a prediction from the lensing theory when the mass distribution exhibits a core structure (\cite{Evans2002, Keeton2003}). Indeed, many of the selected TNG50 subhalos display a core structure near the subhalo centre. However, the presence of such cores could be the result of numerical simulation softening effects. The cores responsible for the central images often emerge within the $3\epsilon^{\text{z=0}}_{\text{DM, stars}}$ boundary, beyond which the gravitational force deviates from Newtonian dynamics due to resolution limitations. Therefore, the presence of these central images and associated cores might not be grounded in physical reality, as they may stem from artificial effects. Such effects have been documented by numerous authors, both in simulations when generating strong lens images from particle data \citep[][]{Mukherjee2018, Enzi2020, He2023, Du2023}, and when investigating other properties of the subhalos \citep[][]{Power2003, Pulsoni2020, Wang2020, WangYunchong2022, Flores-Freitas2022}. On the other hand, on the observational side, the presence of the foreground lens light contamination can lead to the disappearance or challenging detection of central images. To address this issue during our modelling, we follow ref. \cite{He2023} and deliberately increase the assumed flux error in the region containing the central image to such high values that they are effectively ignored during the goodness-of-fit assessment.

A second issue related to the lensing data concerns the boundary truncation effect. This effect arises when the mass map used to create mock observations is improperly truncated, potentially introducing an artificial external shear \citep[][]{Vyvere2020}. The magnitude of this artificial external shear depends on various factors, including the galaxy's profile, truncation area size, and truncation scheme used. In our mock data, we employed a square grid with a length of $80$\,kpc, approximately corresponding to $24$\,arcseconds at $z=0.2$, encompassing roughly $13$ times the typical Einstein radius. However, ref. \cite{Vyvere2020} suggests that the projected convergence map should extend over at least $50$ times the Einstein radius to minimise the impact of this artificial shear. To account for this spurious shear, we will include an external shear component in our modelling, as discussed in section \ref{sec:theory}. We also tested generating mocks using a convergence map with a grid of $480$\,kpc length size, equivalent to about $142$\,arcseconds. This second grid spans approximately $70$ times the typical Einstein radius, aligning with the recommended size outlined by ref. \cite{Vyvere2020}. However, modelling this dataset without a shear component (but with the same model configuration as before) leads to similar results as the ones obtained by the previous dataset.

Our interpretation of these findings is that the artificial external shear appears to enhance the modelling performance, regardless of whether a small or large grid is used for the convergence map. However, it's worth considering that the `artificial' external shear present here might not be as artificial as we initially thought. When ref. \cite{Vyvere2020} created their convergence map, they employed an analytical mass density profile. This method ensures that any external shear in subsequent analyses will be artificial by construction. On the other hand, when using data from hydrodynamical simulations, a larger grid could result in the inclusion of debris from past disruptive interactions or spurious particles at the edges of the subhalo in the convergence map. Given this scenario and our current findings, the external shear contribution we found may indeed be accounting for these spurious contributions.

This interpretation, of course, does not mean that the mass map truncation does not introduce any external shear. What we believe is that a larger convergence map can mitigate the effects of the truncation, but in a scenario where the mock data is derived from a hydrodynamical simulation, this could introduce other sources of contribution to the artificial external shear. For this reason, and keeping in mind that separating this combination could be complex and is beyond the scope of this work, from now on, we will only analyse the results obtained by the dataset generated by the small convergence map and with the inclusion of an external shear. 

Regarding the kinematical data, a significant number of our mock galaxies display a disk-like morphology. This issue stems from a known concern in TNG simulations, where there is known to be an overabundance of red disk-like galaxies \citep[][]{Rodriguez-Gomez2019}. While this does not present a substantial challenge for our purposes, as these disk-like galaxies still fall under the ETG category, we should remain aware of the lack of diversity in our dataset in future investigations.

\section{Methods} \label{sec:theory}
In this section, we briefly introduce the theory and modelling of SGL and stellar dynamics. We also discuss how we combine these different methods self-consistently. 

\subsection{Gravitational lensing}
The SGL phenomena can be described by the lensing equation \cite{Schneider1992},

\begin{equation}\label{eq:lens_equation}
    \bm{\beta}  = \bm{\theta} - \bm{\alpha}(\bm{\theta}),
\end{equation}
where $\bm{\beta}$ is the source-plane position, $\bm{\theta}$ is the image-plane position of the deflected source, and $\bm{\alpha}$ is the reduced deflection angle which maps between the two coordinates, defined as

\begin{equation}\label{eq:reduced_deflection_angle}
	\bm{\alpha}(\bm{\theta}) \equiv  \frac{D_{LS}}{D_S}\bm{\hat{\alpha}}(D_L\bm{\theta}),
\end{equation}
where $\bm{\hat{\alpha}}$ is the deflection angle. The reduced deflection angle depends on the mass distribution of the lens and the angular diameter distances $D_{LS}$, $D_L$ and $D_S$ between the lens and source, lens and observer, and source and observer, respectively. The deflection angle can be obtained from the dimensionless quantity $\kappa$, known as convergence, 

\begin{equation}\label{eq:deflection_angle_convergence}
    \bm{\alpha}(\bm{\theta}) = \frac{1}{\pi}  {\displaystyle \int_{\mathbb{R}^2}} d^2\bm{\theta^\prime} \, \kappa(\bm{\theta^\prime}) \, \frac{\left(\bm{\theta}  - \bm{\theta^\prime}\right)}{\, \, \left|\bm{\theta} - \bm{\theta^\prime}\right|^2},
\end{equation}
where 
\begin{equation}
    \kappa(\bm{\theta}) = \frac{\Sigma(\bm{\theta})}{\Sigma_{\text{cr}}} \quad \text{with } \quad \Sigma_{\text{cr}} \equiv \frac{c^2}{4\pi G}\frac{D_S}{D_L D_{LS}}.
\end{equation}
$\Sigma(\bm{\theta})$ is the surface mass density profile, $c$ the speed of light, and $G$ the gravitational constant. A good approximation for the Einstein radius, $R_{\text{Ein}}$, is the circular radius within which the mean convergence is unity \citep[][]{Tagore2018,Du2023}.

For SGL modelling, we employ the open-source software \texttt{PyAutoLens} \citep{Nightingale2018, PyAutoLens}, which builds upon earlier works by refs. \cite{Warren2003,Dye2008,Nightingale2015}, and implements the Bayesian version of the semi-linear inversion (SLI) method \cite{Suyu2006}, offering substantial enhancements in evaluating the goodness-of-fit in the lens modelling. 

\subsection{Stellar dynamics}
The dynamical state of a collisionless system can be described by the Collisionless Boltzmann equation \citep[][]{Binney_Tremaine2008}. Considering a steady-state axisymmetric configuration, the system should satisfy the two Jeans equations in cylindrical coordinates \cite{Binney_Tremaine2008,Cappellari2008}:

\begin{equation}\label{eq:Jeans}
\begin{aligned}
    \frac{\partial (\nu \overline{v^2_R})}{\partial R} 
    + 
    \frac{\partial (\nu \overline{v_R \, v_z})}{\partial z}
    +
    \nu \left[  \frac{\overline{v^2_R} - \overline{v^2_\phi}}{R} + \frac{\partial \Phi}{\partial R}     \right]
    &=
    0, \\ 
    \frac{1}{R}\frac{\partial \left(R \nu \overline{v_R \, v_z}\right)}{\partial R} 
    + 
    \frac{\partial (\nu \overline{v^2_z})}{\partial z}
    +
    \nu \frac{\partial \Phi}{\partial z}
    &=
    0,
\end{aligned}
\end{equation}
where

\begin{equation}
    \nu \overline{v_i \, v_j} = \int{v_i v_j f d^3 \bm{v}},
\end{equation}
with $f$ being the distribution function of the stars, $\Phi$ the total gravitational potential, $(v_R, v_\phi, v_z)$ the velocities in cylindrical coordinates $(R, \phi, z)$,  and  $\nu$ is the intrinsic luminosity density. 

Following ref. \cite{Cappellari2008}, we additionally assume: (i) the velocity dispersion ellipsoid is aligned with the cylindrical coordinate system (where all the off-diagonal terms vanish); and (ii) a constant flattening of the orbits in the meridional plane, i.e., $\overline{v^2_R} = b \overline{v^2_z}$, where $b$ is the stellar anisotropy. Under these assumptions, the Jeans equations are reduced to:
\begin{equation}\label{eq:Final_Jeans}
\begin{aligned}
    \frac{\partial (b\nu\overline{v^2_z})}{\partial R} 
    + 
    \nu \left[  \frac{b\overline{v^2_z} - \overline{v^2_\phi}}{R} + \frac{\partial \Phi}{\partial R}     \right]
    &=
    0, \\
     \frac{\partial (\nu \overline{v^2_z})}{\partial z}
    +
    \nu \frac{\partial \Phi}{\partial z}
    &=
    0.
\end{aligned}
\end{equation}

Then, imposing the boundary condition $\nu\overline{v^2_z} = 0$  when $z \rightarrow \infty$, the solution can be written as 
\begin{equation}\label{eq:Jeans_v_phi_and_v_z}
\begin{aligned}
    \overline{v^2_\phi} &= b \left[ \frac{R}{\nu} \, \frac{\partial (\nu \, \overline{v^2_z})}{\partial R} + \overline{v^2_z} \right] + R\frac{\partial \Phi}{dR}, \\
    \overline{v^2_z}  &= \frac{1}{\nu}\, {\displaystyle \int_{z}}^\infty d\zeta \, \nu \frac{\partial \Phi}{d\zeta},
\end{aligned}
\end{equation}
where $\zeta$ is the integration variable. These intrinsic quantities should then be integrated along the LOS to obtain the projected second velocity moment $\overline{v^2_{\text{LOS}}}$, directly comparable with the stellar kinematic observable, i.e., $v_{\text{rms}}$.

To solve the axisymmetric Jeans equations and compute the projected second velocity moment, we employ the Jeans Anisotropic Modelling (\cite{Cappellari2008, Cappellari2020}, \texttt{JAM} hereafter) method. The \texttt{JAM} method uses the MGE method to parametrise the mass profile, and given a stellar anisotropy, predict the $v_{\text{rms}}$ by solving the Jeans equations (\ref{eq:Jeans_v_phi_and_v_z}). The goodness-of-fit, in turn, is quantified using a $\chi^2$ statistic to assess the agreement between data and model.

For convenience, the stellar anisotropy parameter in the $z$ direction, $\beta_z$, is rewritten as
$\beta_z = 1 - {\overline{v^2_z}}\, / \, {\overline{v^2_R}} \equiv 1 - 1/b.$

\subsection{Multi-Gaussian Expansion (MGE)}
The mass profile responsible for the stellar orbits and bending the light rays will be parametrised, in this work, as a sum of two-dimensional elliptical concentric Gaussians, the MGE method. Assuming that the stellar mass follows the stellar light profile, the stellar surface brightness can be used as a tracer for the stellar mass density profile. If $I(x^{\prime},y^{\prime})$ is the projected stellar surface brightness, its MGE parametrisation reads

\begin{equation}\label{eq:surf_MGE}
    I(x^{\prime},y^{\prime}) = \sum_{j=1}^{N} \frac{L_j}{2\pi \sigma_{j}^{2} q^{\prime}_j}\exp{\left[-\frac{1}{2\sigma_{j}^{2}} \left(x^{\prime 2} + \frac{y^{\prime 2}}{q_{j}^{\prime 2}}\right)\right]},
\end{equation}
where $N$ is the total number of Gaussians adopted. The $j^{\text{th}}$ Gaussian component has a total luminosity $L_j$, an observed projected axial ratio $0 \leq q^{\prime}_j  \leq 1$, and a dispersion $\sigma_j$ along the semi-major axis, which is aligned with $x^{\prime}$-axis.

The intrinsic three-dimensional luminosity density $\nu$ is obtained by deprojecting equation (\ref{eq:surf_MGE}) assuming an inclination angle $i$, defined as the angle between the shortest axis and the LOS ($i = 90^{\circ}$, when the galaxy is edge-on). The luminosity density is then easily converted to the stellar mass density using a stellar mass-to-light ratio $\Upsilon_\star$. Assuming an oblate axisymmetric model, the stellar mass density profile is written, in cylindrical coordinates, as \citep[][]{Cappellari2002}:

\begin{equation}\label{eq:MGE mass density}
    \rho(R, z) = \sum_{j=1}^N \frac{M_j}{(2\pi)^{3/2} \sigma_{j}^3 q_j} \exp{\left[-\frac{1}{2 \sigma_j^2} \left(R^2 + \frac{z^2}{q_{j}^2}\right)\right]},
\end{equation}
where $M_j = \Upsilon_\star L_j$ is the mass of the $j^{\text{th}}$ Gaussian component with $L_j$ luminosity, $\sigma_j$ are the same as in equation (\ref{eq:surf_MGE}), and $q_j$ is the deprojected three-dimensional intrinsic axial ratio, related to the projected axial ratio by
\begin{equation}\label{eq:_q_deproj}
    q_{j}^2 = \frac{q_{j}^{\prime 2} - \cos^2{i}}{\sin^2{i}}.
\end{equation}

Using the mass density profile, equation (\ref{eq:MGE mass density}), the gravitational potential is obtained using the Homoeoid Theorem for densities stratified on similar concentric ellipsoids \citep[][]{Chandrasekhar1969, Binney_Tremaine2008} as

\begin{equation}\label{eq: Gravitational Potential}
    \Phi(R,z) = -G\sqrt{\frac{2}{\pi}}\sum_{j=1}^{N}\frac{M_j}{\sigma_j}\tilde{\Phi}_j(R,z),
\end{equation}
with $\tilde{\Phi}_j(R,z)$ given by

\begin{equation}    
\tilde{\Phi}_j(R,z) = \int_{0}^{1} \frac{d\tau \exp{\left[-\frac{\tau^2}{2\sigma_j^2} \left(R^2 + \frac{z^2}{1 - \zeta_{j}^{2} \tau^{2}}\right)\right]}}{\sqrt{1 - \zeta_{j}^{2} \tau^{2}}},
\end{equation}
and $\zeta_{j}^{2} = 1 - q_{j}^{2}$.

\subsection{Joint modelling}
To perform the joint analysis and ensure a self-consistent modelling, we combine the phenomena of lensing and stellar dynamics within a Bayesian framework, allowing for the estimation of the posterior distribution of the parameters of interest. Since the lensing and stellar dynamics phenomena are independent, the joint likelihood can be built as the product of the individual likelihoods:

\begin{equation}
    \mathcal{L}_{\text{Model}} = \mathcal{L}_{\text{Lens}} \times \mathcal{L}_{\text{Dyn}}, 
\end{equation}
where  $\mathcal{L}_{\text{Model}}$ is the likelihood of the combined model, $\mathcal{L}_{\text{Lens}}$ is the likelihood of the lens model, and $\mathcal{L}_{\text{Dyn}}$ is the likelihood of the stellar dynamical model. To sample the multi-dimensional parameter space and estimate the posterior distribution of the parameters, we utilised the nested sampler \texttt{dynesty}.

By combining the facilities of \texttt{PyAutoLens}, \texttt{JAM}, and \texttt{dynesty}, we have developed the semi-automatic framework \texttt{dyLens}: dynamical and lens modelling, designed for the self-consistent modelling of SGL and stellar dynamics in galaxies. Within the \texttt{dyLens} framework, the mass profile is represented using the MGE formalism, ensuring simultaneous solution of the Jeans equations (\ref{eq:Jeans_v_phi_and_v_z}), as well as the deflection angle, given by equation (\ref{eq:deflection_angle_convergence}).

\subsection{Pipeline}
For complex and multi-dimensional parameter spaces, the non-linear search may encounter challenges in performing an accurate exploration of the parameter space. Additionally, the SLI method is known to suffer from issues such as under/over-magnified solutions \cite{Maresca2021} for the lens mass model. A way to suppress these erroneous solutions and improve the parameter space sampling is to break the modelling into different phases. Each phase gradually refines the mass model and reduces the parameter space volume, by using the latter phase to adjust the priors of the next one.  In this work, we propose an automatic pipeline similar to those used by different SGL studies (e.g., \cite{Nightingale2018,Cao2022,Etherington2023_shear}). We divide our pipeline into five phases, as described below. 

\textbullet \, \textbf{Phase 1} ({Ph1}): \textit{Parametric Source, Lens + Dynamical modelling} - The modelling is initiated assuming a parametric source profile, for which the under/over-magnified solutions do not exist. In this phase, the parameters describing the mass profile and the source are sampled simultaneously. The source is described by a single Sérsic profile, as given by equation (\ref{eq:sersic}), and the parameters describing the mass profile may vary depending on the model assumptions, i.e, how the stellar mass-to-light ratio is defined or whether a DM halo is included or not. 

\textbullet \, \textbf{Phase 2} ({Ph2}): \textit{Pixelisation} - In this phase, the pixelisation of the source plane is initialised. The parameters describing the mass profile model are fixed based on the results of Ph1 and a non-linear search is conducted to determine the hyper-parameters of the source plane. We assume a Delaunay tessellation with a constant regularisation scheme in \texttt{PyAutoLens}.  The Delaunay tessellation uses an irregular mesh of Delaunay triangle pixels, adapting to the mass-model magnification pattern and placing more source pixels in highly magnified regions of the source plane, while the constant regularisation term is responsible for penalising sources solutions that are less smooth (see, e.g., refs. \cite{Warren2003, Dye2008}).

\textbullet  \, \textbf{Phase 3} ({Ph3}): \textit{Model Refinement} I -  In this phase, the mass model parameters are refined, now using a pixelated source plane. The source plane parameters are fixed using the information of Ph2, and the prior knowledge of the mass model parameters is updated accordingly to the Ph1 results. This approach not only reduces the parameter space to mitigate the under/over-magnified solutions but also refines the model, enabling the true morphology of the source to be revealed. 

\textbullet \, \textbf{Phase 4} ({Ph4}): \textit{Adaptive Brightness-based Pixelisation and Regularisation} - In this phase, the source plane pixelisation and regularisation are enhanced to adapt to the source surface brightness of the lensed source, instead of the mass model. The adaptive brightness-based pixelisation ensures that areas with higher fluxes are reconstructed with higher resolution, while areas with lower fluxes are reconstructed with fewer pixels \citep[][]{Nightingale2018}. Similarly, the adaptive brightness-based regularisation applies non-constant regularisation to each pixel, heavily penalising areas with lower flux. We assume the  {\fontfamily{qcr}\selectfont DelaunayBrightnessImage} pixelisation with a {\fontfamily{qcr}\selectfont AdaptiveBrightnessSplit} regularisation in \texttt{PyAutoLens} \footnote{See the online material for further information: \url{https://pyautolens.readthedocs.io/en/latest/api/pixelization.html}}.

\textbullet \, \textbf{Phase 5} ({Ph5}): \textit{Model Refinement} II - The last phase of the pipeline corresponds to a fine refinement of the mass model. To achieve this and avoid the issue of small statistical uncertainties identified by ref. \cite{Etheringto2022}, we first compute the likelihood cap as described and proposed by them. The parameters used to compute the cap are fixed using the Ph3 (mass model) and Ph4 (source-plane) results, and we use $300$ likelihood evaluations. After that, with the source plane parameters fixed, a new nested sampling run for the mass model parameters is performed, for which the priors are updated using the Ph3 information. Ph5 model is considered the ``final'' model and will be used for further analysis.

It is important to highlight that this pipeline is only applied when lensing data is being modelled, since the dynamical modelling does not suffer from the under/over-magnified solutions. Additionally, to improve the computational speed, we applied a circular mass, centred at the image centre, to model only the inner $2.5$\,arcsec pixels.

\subsubsection{Prior update}
To update the priors, we use the median value of the one-dimensional marginalised posterior distribution of each parameter. These median values are also used when the model parameters are fixed, e.g., when in Ph3 the source plane parameters are fixed. The priors are updated while maintaining their form but reducing the parameter space volume.

If $\mathbf{\hat{\Theta}}_\mathbf{U}$ corresponds to the median values of a sub-set of parameters that have uniform/log-uniform priors, the new bounds are set to $\mathbf{\hat{\Theta}}_\mathbf{U} \pm \alpha_p |\mathbf{\hat{\Theta}}_\mathbf{U}|$ or $\mathbf{\hat{\Theta}}_\mathbf{U} \pm \max|1\sigma(\mathbf{\hat{\Theta}}_\mathbf{U})|$, whichever yields a larger interval. Here, $1\sigma(\mathbf{\hat{\Theta}}_\mathbf{U})$ denotes the parameter uncertainty, measured as the $68\%$ credible intervals and obtained by the $16$th and $84$th percentiles, and $\alpha_p \in [0,1]$. The updated priors are accepted if the new bounds are smaller; otherwise, the original priors remain.  For parameters with a normal distribution prior, the mean and the dispersion are both updated. The new mean is set as the median value of the one-dimensional posterior distribution of the associated parameter, while the new dispersion is defined as $\alpha_p |\mathbf{\hat{\Theta}}_\mathcal{N}|$ or $\max|1\sigma(\mathbf{\hat{\Theta}}_\mathcal{N})|$, whichever is large. The updated dispersion is accepted if smaller; otherwise, the original is kept.

In this work, we assume $\alpha_p = 0.2$ for Ph3 and $\alpha_p = 0.1$ for Ph5.

\subsection{Mass model}\label{sub:mass_profile}
We model the mock galaxies using the same mass model and the pipeline outlined earlier. The mass profile is separated into a stellar mass contribution and a DM halo. The stellar mass distribution is obtained from the MGE surface brightness profile, which is derived from the projection of the stellar particles (see Sec. \ref{sec:sample}). The DM halo is represented by a gNFW profile, as given by equation (\ref{eq:gNFW}). We assume a constant mass-to-light ratio and a constant stellar anisotropy, consistent with the mock preparation. To address the issue of boundary truncation, as presented by ref. \cite{Vyvere2020} and discussed in section \ref{sec:sample}, we also include an external shear component. We applied this same mass profile to three modelling strategies: dynamical-only\footnote{Note that the external shear component only appears when lensing data is modelled.} (Dyn), lens-only (Lens), and joint (dyLens).

More details about the mass parameters and the applied priors are given in Appendix \ref{ap:mass_model}.

\section{Results and Discussion}\label{sec:results}
In this section, we present and discuss our results regarding the modelling of the simulated sample using the Dyn model, Lens model and dyLens model. In section \ref{sub:fiducial_conf} we show the results obtained using the mass profile described in section \ref{sub:mass_profile} and using the kinematical data with 35 bins (see section \ref{sec:sample} for details). We will refer to this case as the fiducial configuration. Then, in section \ref{sub:kin_bias}, we present the results obtained when the number of kinematical tracers is changed. Finally, in section \ref{sub:mass_bias}, we discuss the implication of imposing more restrictive assumptions in the DM profile.

\subsection{Accuracy in recovered mass model}

We assess the accuracy and effectiveness of the models in recovering various quantities, including intrinsic and projected stellar and DM mass, DM fraction, and total mass. We also investigate if the models can constrain specific parameters like those in the gNFW profile and the stellar mass-to-light ratio. Additionally, we explore how the Dyn and dyLens models recover the total mass density slope $\gamma$.

To measure the intrinsic and projected quantities within a simulated galaxy, we assume spherical symmetry for intrinsic values and cylindrical symmetry for projected values. We consider the contributions of all particles within a specified radius. The model's intrinsic mass is estimated using the analytic mass of an axisymmetric MGE within a sphere of a given radius \citep[][]{Mitzkus2017}:

\begin{equation}
    M(r)_\text{model} = \sum_j M_j\left[ \text{erf}(h_j) -\frac{\exp{(-h^2_jq^2_j)}\text{erf}(h_je_j)}{e_j}   \right],
\end{equation}
with
\begin{equation}
     \begin{aligned}
        e_j = \sqrt{1 - q^2_j}, \quad
        h_j = \frac{r}{\sigma_j q_j \sqrt{2}},
    \end{aligned}
\end{equation}
where $\text{erf}(x)$ is the error function and $j$ runs over all the Gaussian components of the MGE mass model\footnote{We use the {\fontfamily{qcr}\selectfont mge\_radial\_mass} implemented in \texttt{Jampy}.}. 

The projected mass model is obtained by integrating the MGE surface mass profile $\Sigma(x^\prime,y^\prime)$ within a circular radius $R$:
\begin{equation}
    M(R)_\text{model} = 2\pi \int_0^R dR^\prime \, R^\prime  \, \Sigma(R^\prime) = \sum_j M_j \left[ 1 - \exp{\left( \frac{-R^2}{2\sigma^2_j} \right)} \right],
\end{equation}
where $j$ runs over all the Gaussian components of the MGE mass model.

To estimate the total mass density slope we use two different approaches since there is no consensus in the literature on the best way to measure it (see refs. \cite{Dutton2014,Li2016,Poci2017,Bellstedt2018,Derkenne2023}). First, we compute the average slope $\gamma_{\text{AV}}(r_1 ,r_2)$ of the density profile between two radii $r_1$ and $r_2$, as suggested by ref. \cite{Xu2017} and defined as

\begin{equation}\label{eq:average_slope}
    \gamma_{\text{AV}}(r_1 ,r_2) = \frac{\ln{\left[\rho(r_2)/\rho(r_1)\right]}}{\ln{\left(r_1/r_2\right)}},
\end{equation}
where $\rho(r)$ is the total density profile from the simulated galaxy or from the model. To ensure consistency, we sample the model's total density profile\footnote{We use the {\fontfamily{qcr}\selectfont mge\_radial\_density} function implemented in \texttt{Jampy} to calculate the spherically-averaged density of an axisymmetric MGE model, which is given by equation (11) of ref. \cite{Cappellari2015-density}.} in the same $100$ radial bins as those used in the density profile for the simulations. Then, to obtain the value of the density profile at the specific radii $r_1$ and $r_2$, we interpolate across the radius using a cubic spline interpolator. 

The second estimation of the density slope is obtained by fitting a power-law profile to the total density profile between $r_1$ and $r_2$ \citep[][]{Remus2013,Li2016}. We will refer to this estimation as $\gamma_{\text{PL}}(r_1 ,r_2)$.

For models able to constrain the Einstein radius $R_{\text{Ein}}$, we are also interested in determining their accuracy. We measure $R_{\text{Ein}}$ as the circular radius within which the mean convergence is unity \citep[][]{Tagore2018,Du2023}. The reference value is directly computed from the convergence used to generate the mock lenses, while the modelled value is obtained from the model's convergence. 

To compute all these quantities, we utilised two approaches. Firstly, we considered the median value of the one-dimensional posterior distribution of parameters obtained after Ph5. This approach is computationally efficient and suitable for establishing a fiducial model configuration. However, it does not fully consider parameter covariance, and error propagation is not straightforward. Secondly, we reconstructed the posterior distribution for each desired quantity using all parameter sets in the nested run. This allows us to obtain the posterior distribution for the quantity of interest, from which the median value can be extracted as a representative model outcome. While this method is computationally demanding, it captures the full parameter covariance from the non-linear run and provides a distribution for the desired quantity, allowing uncertainty to be calculated through credible intervals. We advance that, based on our analysis, both methods yield comparable results. Given that the second approach provides a more robust estimate of uncertainties, we will only present and discuss the findings obtained by this method.

To assess the accuracy in a quantity $Q$, we used the fractional error defined by
\begin{equation}
    \Delta Q = \frac{Q_\text{model} - Q_\text{data}}{Q_\text{data}},
\end{equation}
where $Q_\text{model}$ is the estimated quantity derived from the model output, and $Q_\text{data}$ is the same quantity extracted directly from the simulated galaxy or the ``true'' input.

\subsection{Fiducial configuration}\label{sub:fiducial_conf}

In this section, Dyn results will consistently be represented by dashed black lines, Lens results by dash-dotted red lines, and dyLens results by hatched blue lines. Furthermore, each panel maintains the same layout unless stated otherwise: upper left for stellar mass, upper right for DM mass, lower left for DM fraction, and lower right for total mass. We will make it clear throughout the text whether the quantities discussed are intrinsic or projected, along with the corresponding radius at which these quantities were computed.

\subsubsection{Enclosed quantities}
\begin{figure}[h!]
	\includegraphics[width=\columnwidth]{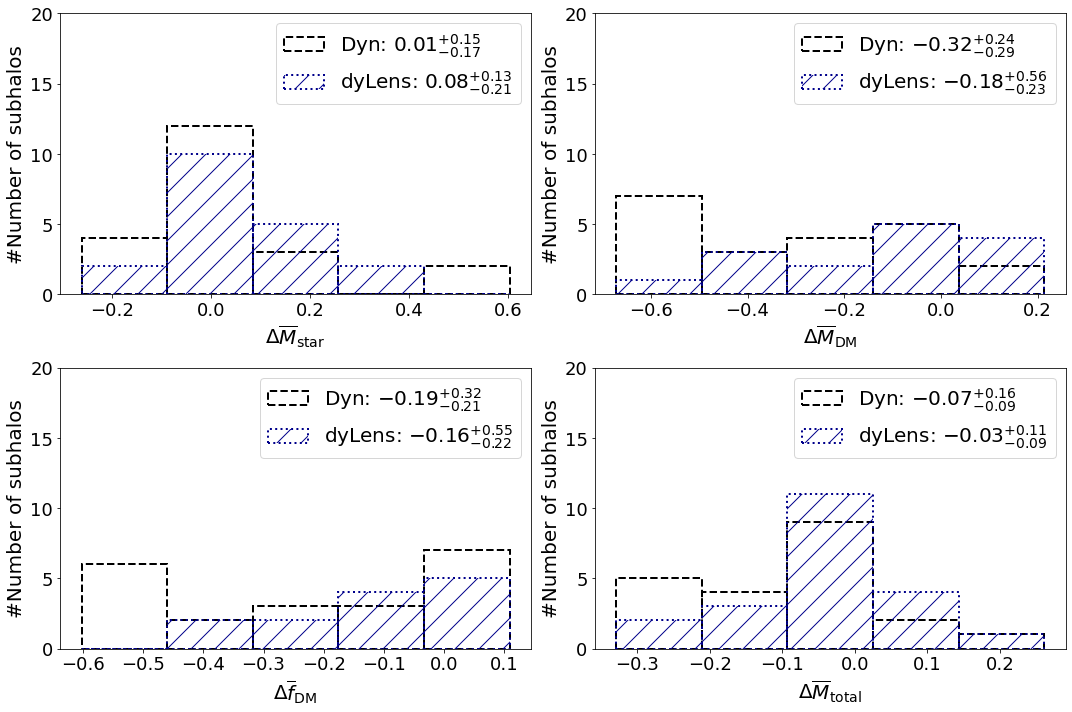}
    \caption{Fractional error in 3D intrinsic enclosed quantities within $2.5R_\text{eff}$. From upper left to lower right: stellar mass, DM mass, DM fraction and total mass. The inset legend in each panel shows the median and the $68\%$ spread around it. Dyn results are shown in dashed black and dyLens results in hatched blue.}
    \label{fig:bias_3D_distribution_normalFOV}
\end{figure}

Figure \ref{fig:bias_3D_distribution_normalFOV} show the fractional errors in the intrinsic enclosed quantities within $2.5R_\text{eff}$. Overall, the dyLens model presents a smaller median fractional error for all quantities except stellar mass, where the Dyn model performs better ($1\%$ for Dyn, against the $8\%$ for the dyLens). The enclosed DM mass exhibits a significant improvement with the joint model when compared with the Dyn model (from a median value of $-32\%$ with Dyn to a $-18\%$ with dyLens). However, it is interesting to note that the spreads around the median values in the DM components are significantly large for both models, which can make conclusions about individual galaxies challenging to obtain.

The median fractional error in the recovered enclosed total mass for the dyLens model is only $-3\%$, an improvement compared to the $-7\%$ in the Dyn model. This good recovery is further supported by the small scatter around the medians, which is $9-11\%$ for dyLens and $9-16\%$ for Dyn. The fact that the enclosed total mass is better recovered than the other quantities is not surprising, once both phenomena, stellar dynamics and SGL, are only sensible to the total gravitational potential $\Phi_\text{total}$, which is determined by both the DM and the stellar masses. 

While a good recovery of the enclosed total mass is expected, it is noteworthy that the recovery of the enclosed stellar mass is much better than that of the enclosed DM mass, as seen in Figure \ref{fig:bias_3D_distribution_normalFOV} (both upper panels). Furthermore, some outliers with $\Delta \overline{M}_\text{DM} \gtrsim -0.6$ and $\Delta \overline{M}_\text{star} \sim 0.6$ for the dyLens model (but also present in the Dyn model) suggest a correlation between these errors. This correlation was previously observed by ref. \cite{Li2016} when evaluating the Jeans Anisotropic Modelling. Ref. \cite{Li2016} pointed out that the stellar mass component is measured from the MGE decomposition of the observed image, where only the mass-to-light ratio is allowed to vary. Therefore, limited image resolution in MGE modelling can lead to a bias in disentangling the DM and the stellar contributions. Additionally, the DM mass component has more degrees of freedom and is not directly probed, further complicating its constraint. Consequently, we suspect that even with better image resolution this correlation may persist, once the typical galaxy SGL phenomenon and galaxy stellar kinematics dataset can only constrain the mass profile in the central regions, where the DM parameters are challenging to constrain, especially the scale radius.

When evaluating the same quantities within the Einstein radius, which is typically smaller than $2.5R_\text{eff}$, we find a median fractional error of $3\%$ in the recovered stellar mass for the dyLens model\footnote{For consistency, we do not measure any quantities within $R_\text{Ein}$ for the Dyn model, once stellar dynamics-only is not able to measure $R_\text{Ein}$.}, while the median fractional error for the recovered DM mass is much worse, at $-39\%$. However, the recovering of the total mass is consistent with that found within $2.5R_\text{eff}$, i.e., $ \langle \Delta \overline{M}_\text{total}(R_\text{Ein}) \rangle = -0.03^{+0.13}_{-0.09}$, which again may reinforce the idea that the DM profile is weakly constrained in the innermost regions of the galaxy.

\begin{figure}[h!]
	\includegraphics[width=\columnwidth]{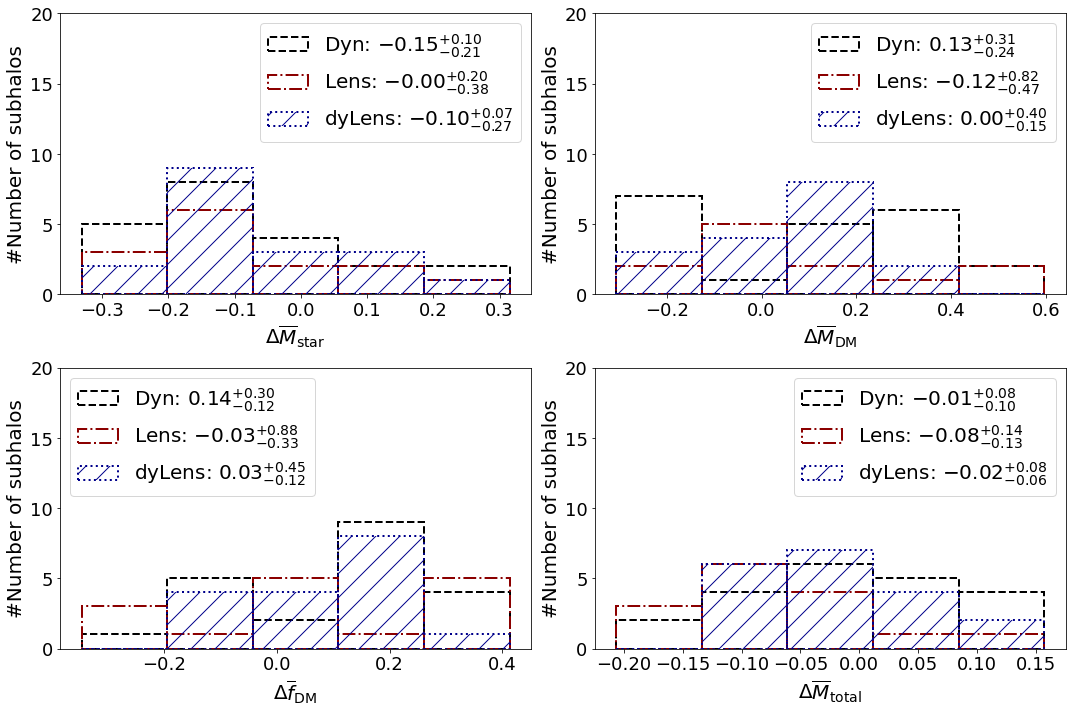}
    \caption{Fractional error in the 2D enclosed quantities within $2.5R_\text{eff}$. From upper left to lower right: stellar mass, DM mass, DM fraction and total mass. The inset legend in each panel shows the median and the $68\%$ spread around it. Dyn results are shown in dashed black, Lens in dash-dotted red and dyLens results in hatched blue.}
    \label{fig:bias_2D_distribution_normalFOV}
\end{figure}

In Figure \ref{fig:bias_2D_distribution_normalFOV} we show the fractional error in the projected enclosed quantities within $2.5R_\text{eff}$ for the three models. The dyLens model outperforms the Dyn model for all quantities except the enclosed total mass, where Dyn performs slightly better. The Lens model yields better results than the Dyn model for the enclosed stellar mass and DM fraction, with a similar recovery for the DM mass. It is worth noting that the spread around the median in the Lens model is significantly larger than in the other models.

In terms of enclosed total mass, both Dyn and dyLens yield robust and nearly unbiased estimates, displaying only a minor spread in their results. The Lens results exhibit a slightly higher median fractional error of $-8\%$, along with a wider spread. As seem in previous results, some outliers are present, with errors surpassing $60\%$ in the recovered enclosed DM mass. Notably, one subhalo (Id. 7) shows $\Delta \overline{M}_\text{DM}(2.5R_\text{eff}) = 1.58$ and $\Delta \overline{M}_\text{total}(2.5R_\text{eff}) = 0.67$, where visual inspection confirms a poor fit of the lensed image.

\begin{figure}[h!]
	\includegraphics[width=\columnwidth]{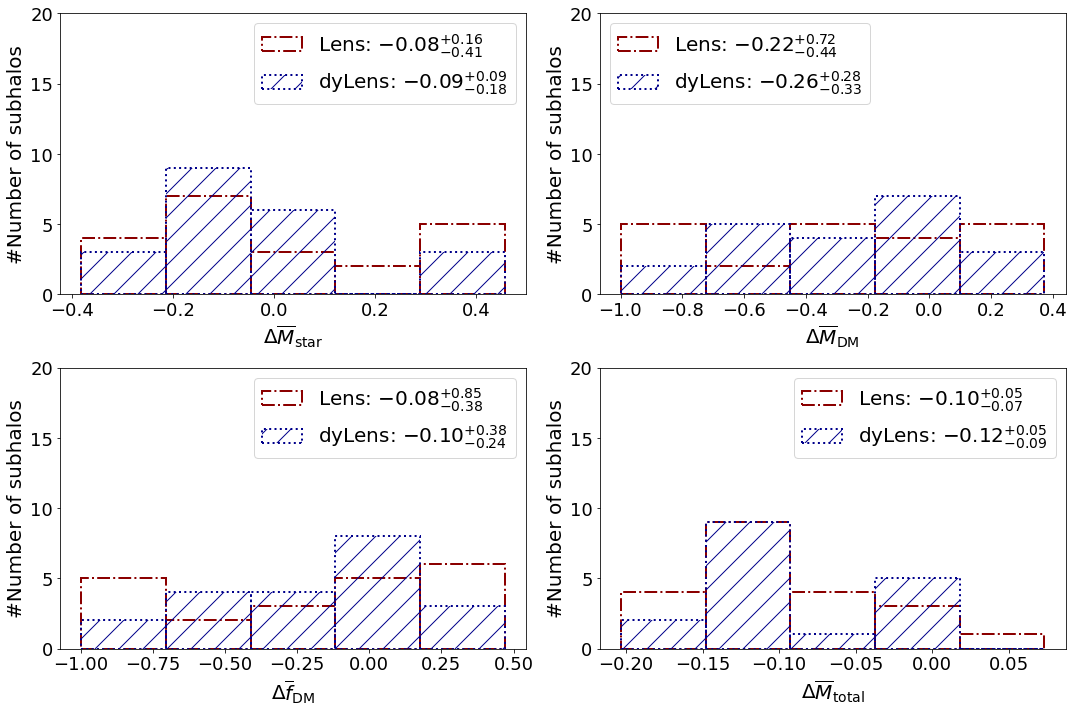}
    \caption{Fractional error in the 2D enclosed quantities within $R_\text{Ein}$. The panels and colour scheme are the same as in Figure \ref{fig:bias_2D_distribution_normalFOV}.}
    \label{fig:bias_2D_distribution_normalFOV_Re}
\end{figure}

Examining projected quantities within $R_\text{Ein}$ reveals an unexpected behaviour. As shown in Figure \ref{fig:bias_2D_distribution_normalFOV_Re}, the median fractional error in the Lens model slightly outperforms the dyLens model across all quantities. However, the spread around the median is notably wider for the Lens model. This discrepancy may be explained by Figure \ref{fig:bias_2D_distribution_normalFOV}, where it is evident that the dynamical-only model consistently underperforms compared to other models in recovering projected quantities. This limitation could potentially affect the recovery of the dyLens model. Also note that the Lens model is only marginally better than the dyLens model, with an improvement of only a few per cent (e.g., $4\%$ in the best case for the enclosed DM mass), but at the cost of a larger scatter. This suggests that the dyLens model remains competitive in this context.

\subsubsection{Total density slope}

\begin{figure}[h!]
	\includegraphics[width=\columnwidth]{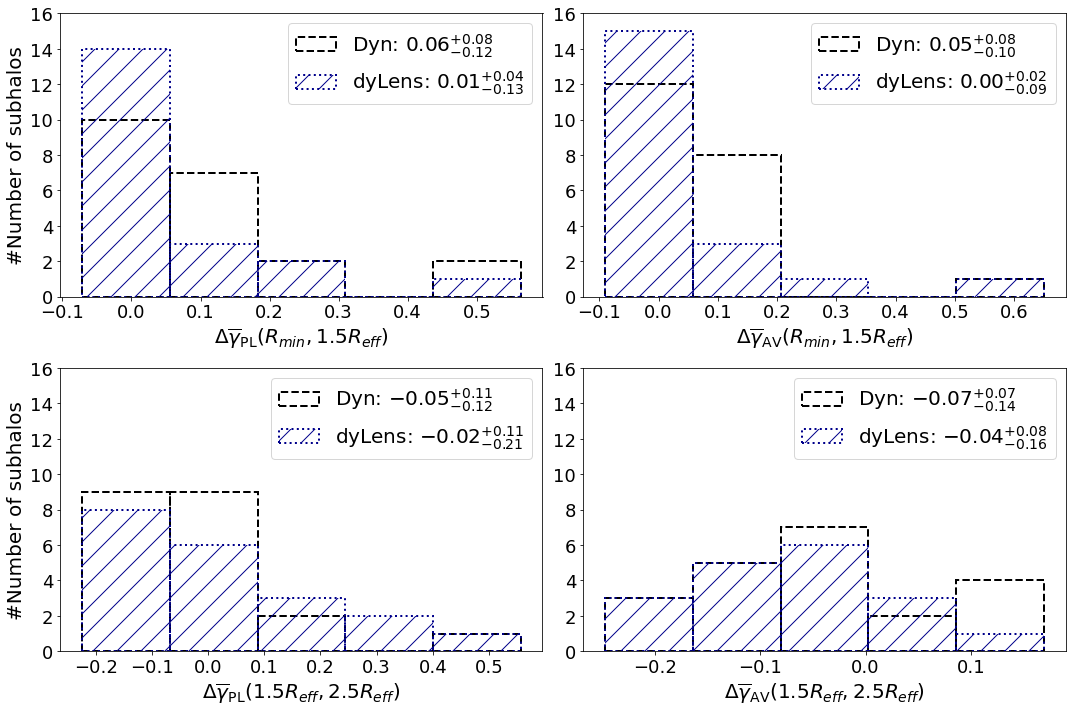}
    \caption{Fractional error in the recovered total density slope. The upper panels are the slopes measured between $r_\text{min}$ and $1.5R_\text{eff}$, while the bottom panels cover the range from $1.5R_\text{eff}$ to $2.5R_\text{eff}$. The left column represents the power-law slope and the right column the average slope. The inset legend in each panel shows the median and the $68\%$ spread around it. Dyn results are shown in dashed black and dyLens results in hatched blue.}
    \label{fig:slope_distribution}
\end{figure}

Figure \ref{fig:slope_distribution} displays the fractional error in the recovered total density slope derived from the density profile obtained from the models\footnote{We used the \texttt{scipy} routine \texttt{mquantiles} to compute the median curve ($50\%$ quantile) from which the slopes were measured.}. The upper panels show the power-law (left) and average (right) fractional errors in the total density slopes within the region between $r_\text{min}$ and $1.5R_\text{eff}$ for both the Dyn and dyLens models, where $r_\text{min}$ is the minimum radial bin of the mass density profile extracted from the simulation. The Dyn model exhibits a small median fractional error of $6\%$ for the power-law slope and $5\%$ for the average slope. For the dyLens model, the average and power-law slopes show a median bias consistent with a zero and a typical spread of $\sim 10\%$. The bottom panels show the same quantities as above, but they are measured within the region between $1.5R_\text{eff}$ and $2.5R_\text{eff}$. Here, we observe that the median fractional error shows a small increase for all models, whether considering the power-law (left) or average (right) slopes. However, all median values remain below a $10\%$ bias. Noteworthy, all models tend to underestimate the total density slope\footnote{There is an exception in the measured slopes for subhalo116278. For this subhalo, we measured the total mass density slopes within the apertures ($r_\text{min}$, $2R_\text{eff}$) and ($2R_\text{eff}$, $2.5R_\text{eff}$) since its effective radius is too small and causes numerical issues during the interpolation.}.

\subsubsection{Parameter estimation}

In studies related to DM, we might be interested in the parameters that describe the DM profile, such as the inner slope, which can help differentiate between cusp and core halos \citep[][]{Blok2010, Popolo2022}. Similarly, in the galaxy evolution and stellar population analysis, estimates of the mass-to-light ratio are crucial for characterising the initial mass function \citep[][]{Sonnenfeld2012,Cappellari2012_IMF}.

To compare the DM parameters, we treat the median of the one-dimensional posterior distribution obtained from the direct non-linear fit of the radial profiles (see Sec. \ref{sec:sample}) as the ``true'' value, however, we will still report the uncertainty. For the model outcome, we will consider the median parameter value obtained from the one-dimensional posterior distribution of Ph5. For both median values, we will consider the $68\%$ credible interval as the $1\sigma$ uncertainty.

\begin{figure}[h!]
	\includegraphics[width=\columnwidth]{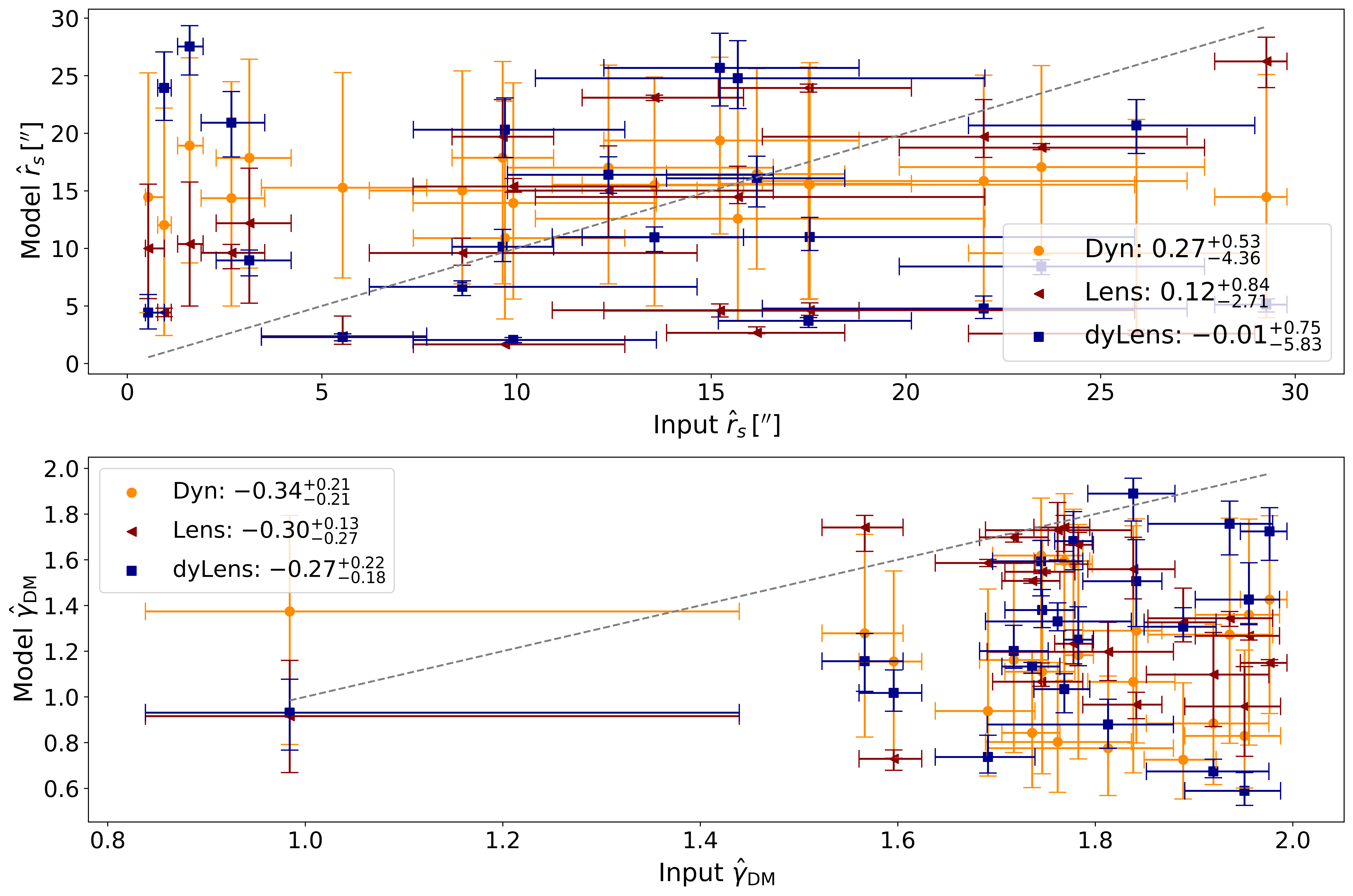}
    \caption{Recovering of the DM scale radius (upper) and DM inner slope (lower). Dyn model is shown in orange dots, Lens in red triangles and dyLens in blue squares. The error bars are the $68\%$ credible levels from the model or the direct fit. The inset panels show the median fractional error and the $68\%$ spread around it. The dashed line represents the identity curve to guide the eye.}
    \label{fig:dm_parameters_recover}
\end{figure}

Figure \ref{fig:dm_parameters_recover} illustrates the relation between the ``true'' and model scale radius (upper panel) and DM inner slope (lower panel) for the $21$ systems in our sample.  It is evident that neither parameter is well-constrained and the error bars are quite large, even for the ground base values that were directly fitted to the DM density profile. However, it is surprising that the dyLens model recovers the scale radius with a very small median fractional error, albeit with a trade-off of a large scatter.

Furthermore, it is worth noting that across nearly all mock galaxies, there is a consistent underestimation of the DM inner slope. Specifically, all models display a median fractional error of approximately $-30\%$. However, as previously observed, the total density slope remains well-constrained. Thus, we attribute these discrepancies in the recovered DM parameters to two primary factors: parameter degeneracy and the extent of spatial data coverage, which pose challenges in constraining DM parameters accurately.

Regarding the stellar mass-to-light ratio, we found that the Dyn model performs extremely well, with a median fraction error of $+0.01^{+0.08}_{-0.19}$ and recovering, for almost all subhalos, the input value within $1\sigma$. The Lens and dyLens models show a good recovery (although worse than the dynamical-only model), with median fractional errors of $+0.15^{+0.24}_{-0.29}$ and $+0.11^{+0.16}_{-0.24}$, respectively. Albeit the Lens and dyLens have a similar median fractional error, the Lens model has a larger scatter. The fact that the Dyn model outperforms the other two models is not surprising, in light of the previous results, where the Dyn model tends to recover the total stellar mass more accurately compared to the other models.

As discussed earlier, only the total mass profile is constrained by the data, which makes it difficult to separate the stellar and DM components directly. One way to distinguish between them is based on their inner mass density slope \citep[][]{Sonnenfeld2013,Jin2019}. However, the shape of the stellar density profile is fixed by the MGE parametrisation, with only the overall amplitude allowed to vary through changes in the mass-to-light ratio. Additionally, as observed in Figure \ref{fig:dm_parameters_recover}, the DM inner slope is not well constrained, implying some correlation between the fractional errors in the mass-to-light ratio and the DM inner slope. Something similar was first reported by ref. \cite{Li2016} (in their Figure 6), where there is a clear correlation between the fractional error in the enclosed stellar and DM masses.

To compute the model Einstein radius, we first create a model convergence map using the median value of the one-dimensional posterior distribution of mass parameters obtained during Ph5. The dyLens model exhibits a median fractional error of $+0.00^{+0.26}_{-0.03}$, while the Lens model recovers $R_\text{Ein}$ with a median of $-0.05^{+0.15}_{-0.05}$. When compared to the power-law model studied by ref. \cite{Cao2022}, which achieves an accuracy of $0.1\%$ in recovering the Einstein radius, our dyLens model shows a slight improvement, while the Lens model is competitive.

In the dyLens model, we have found a notable outlier with $\Delta \hat{R}_\text{Ein} < -1.0$, which corresponds to subhalo451938, one of the less massive objects in our sample. We checked the fitted model and the source reconstruction, and after that, we confirmed that it resulted in a poor fit for both. Interestingly, the integrated quantities within $2.5R_\text{eff}$ are well constrained despite this poor fit, with the intrinsic total enclosed mass recovered with $9\%$ accuracy. However, the integrated quantities within the Einstein radius are less well-constrained, with the intrinsic total enclosed mass recovered with $15\%$ accuracy. This suggests that the inner mass distribution for this specific subhalo has not been fully captured, although the overall mass distribution was captured, at least for $R < 2.5R_\text{eff}$. If we choose to exclude this subhalo from the analysis, the upper limit of the spread around the median fractional error decreases from $0.26$ to $0.17$, which represents a significant improvement.

\subsection{Impact of the number of kinematical constraints}\label{sub:kin_bias}

Here we investigate how the number of bins in the kinematical map impacts the constraining power of the models. We consider two variations in the number of bins, as present in section \ref{sec:sample}. One with $\sim15$ bins and another with  $\sim55$ bins, in addition to the previous dataset containing $\sim35$ bins. All other inputs remain consistent with our former analyses, and we ran the Dyn and dyLens models using these new kinematical data configurations. We will use the following notation hereafter: Model1 refers to the fiducial model with $\sim35$ bins, Model2 is the model with $\sim15$ bins and Model3 represents the model with the larger number of kinematical bins.

\begin{figure}[h!]
	\includegraphics[width=\columnwidth]{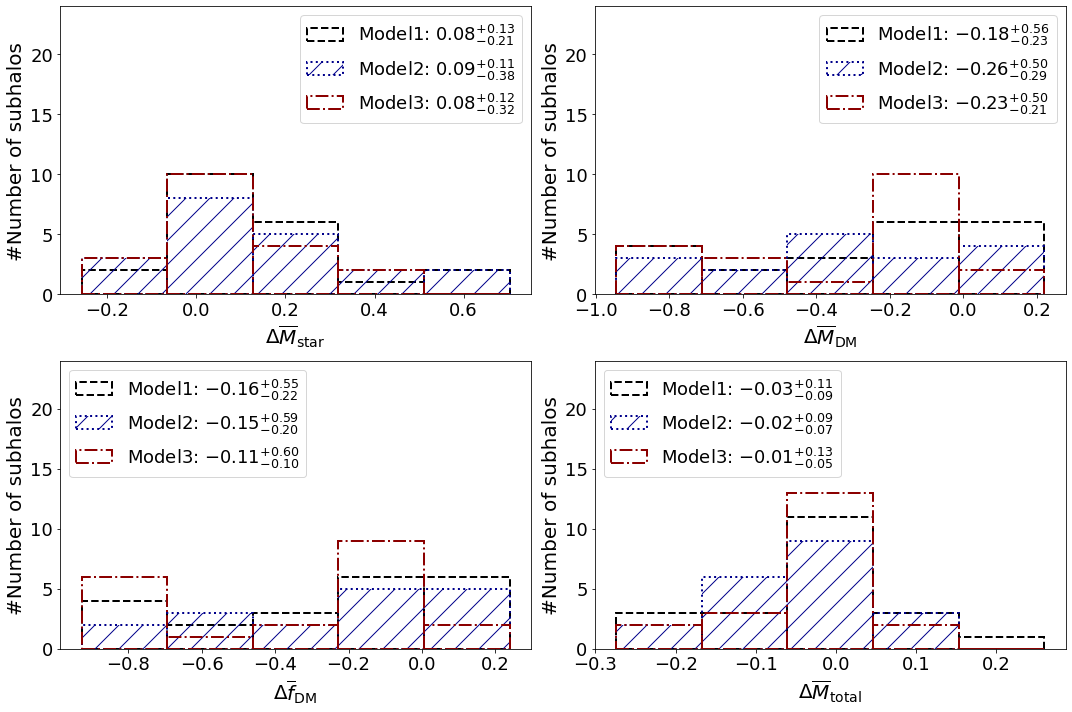}
    \caption{Fractional error in the recovered 3D enclosed quantities within $2.5R_\text{eff}$ for dyLens models considering different number of kinematical constraints. Model1 ($\sim15$) is represented in dashed black, Model2 ($\sim35$) in hatched blue and Model3 ($\sim55$) in dash-dotted red. From upper left to lower right: stellar mass, DM mass, DM fraction and total mass. The inset legend in each panel shows the median and the $68\%$ spread around it. }
    \label{fig:bins_comparison_dyLens_3D}
\end{figure}

In Figure \ref{fig:bins_comparison_dyLens_3D}, we present the fractional error in the recovered intrinsic properties for the dyLens models within $2.5R_\text{eff}$. It is possible to see that the number of bins in the kinematical map has minimal to no impact on the medians of the intrinsic enclosed quantities. The results for the projected quantities obtained from the dyLens models are similar, with the number of kinematical bins showing no impact. This conclusion also extends to measurements carried out within $R_\text{Ein}$.

A similar conclusion is reached when considering the results (intrinsic and projected quantities) of the Dyn models, i.e., the number of kinematical constraints does not affect the model outcomes, considering the number of constraints employed in this work. 

We have also checked that the total mass density slopes and the parameter estimations are not significantly affected by the number of kinematical bins, for both Dyn and dyLens models. Not only the median fractional errors are consistent, so are the spreads.

\subsection{Impact of more restricted assumptions about the DM halo}\label{sub:mass_bias}

As demonstrated by Figure \ref{fig:dm_parameters_recover}, the DM parameters are typically not well-constrained by the datasets. These weak constraints may be affecting the recovery of the integrated DM properties. The primary factor we hypothesise to be responsible for these weak constraints is the spatial extent of the data. Essentially, both datasets cannot effectively constrain the scale radius of the DM profile. Thus, keeping this parameter as a free variable introduces degeneracies that the data cannot overcome. In this context, imposing more restrictive assumptions about the DM profile could be useful in alleviating such issues. One common approach, motivated by numerical simulations (see ref. \cite{Kravtsov2013}), is to fix the scale radius as being proportional to the stellar light effective radius (e.g., \cite{Sonnenfeld2015,Collett2018}). We explored this assumption and considered a DM mass profile with the scale radius fixed at $10$ times the stellar effective radius, i.e., $r_s = 10R_\text{eff}$. We ran the Dyn and dyLens models assuming the same dataset and priors of Model1 ($\sim 35$ kinematical bins) while keeping all the inputs consistent with our previous analyses.

It is important to acknowledge that this assumption about the DM scale radius does not hold for almost all of our mock datasets, with $r_s$ being typically greater than the $R_{\text{eff}}$. Therefore, this assumption should be interpreted as a proxy and a means to reduce degeneracies rather than as a ``true'' estimate for the DM scale radius.

\begin{figure}[h!]
	\includegraphics[width=\columnwidth]{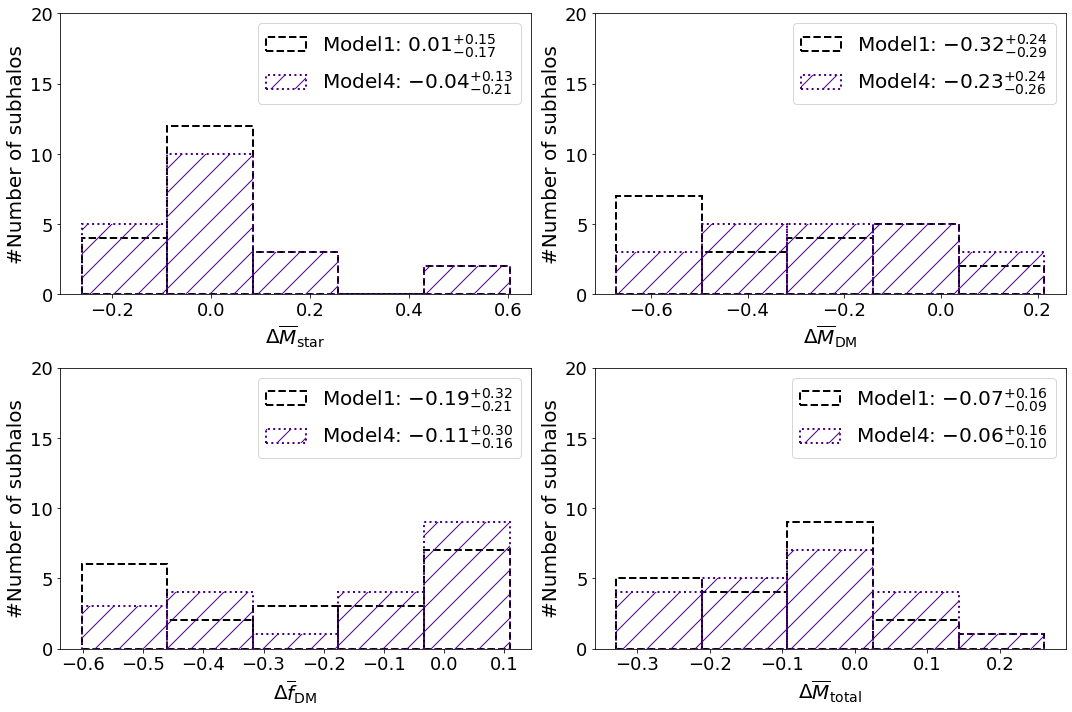}
    \caption{Fractional error in the recovered 3D enclosed quantities within $2.5R_\text{eff}$ for Dyn models. From upper left to lower right: stellar mass, DM mass, DM fraction and total mass. The inset legend in each panel shows the median and the $68\%$ spread around it. Model1 (varying $r_s$) is represented in dashed black and Model4 (fixed $r_s$) is in hatched purple.}
    \label{fig:DM_comparison_dynamics_3D}
\end{figure}

Hereafter, we will refer to the model with the DM scale radius fixed as Model4, and compare it with Model1. Figure \ref{fig:DM_comparison_dynamics_3D} shows the recovery of the intrinsic properties within $2.5R_\text{eff}$ obtained from dynamical-only modelling for Model1 and Model4. A significant improvement (around $10\%$) is seen in the median error of the enclosed DM mass and DM fraction, when the DM scale radius is fixed. However, the spread is still similar for both models.

\begin{figure}[h!]
	\includegraphics[width=\columnwidth]{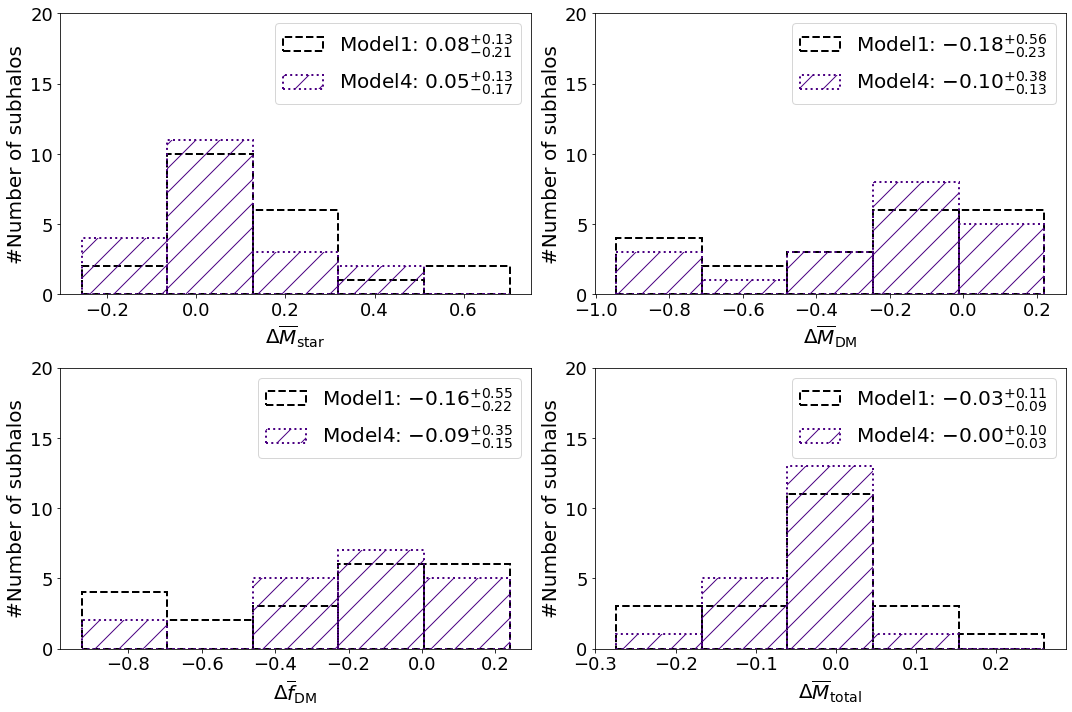}
    \caption{Fractional error in the recovered 3D enclosed quantities within $2.5R_\text{eff}$ for dyLens models. The panels and colour scheme are the same as in Figure \ref{fig:DM_comparison_dynamics_3D}.}
    \label{fig:DM_comparison_dyLens_3D}
\end{figure}

The dyLens models are considered in Figure \ref{fig:DM_comparison_dyLens_3D}. In this case, it is possible to see an improvement in all the quantities being considered, with the major improvements being in the enclosed DM mass and DM fraction. It is also worth noting that for these two quantities, the spread around the median is considerably reduced for Model4. A similar trend is observed for the intrinsic recovery quantities within $R_\text{Ein}$, showing that Model4 outperforms Model1 in recovering the intrinsic enclosed DM mass and intrinsic DM fraction.

When considering the projected quantities within $2.5R_\text{eff}$, the Dyn models perform with a similar median fractional error for all the quantities but the projected enclosed DM mass. The latter is recovered for Model1 with $\langle\Delta \overline{M}_\text{DM} \rangle = 0.13^{+0.31}_{-0.24}$, while for Model4 $\langle\Delta \overline{M}_\text{DM} \rangle = 0.02^{+0.30}_{-0.23}$, showing a considerable improvement, despite the huge spread in both cases.

Regarding the dyLens models, the projected quantities within $2.5R_\text{eff}$ are recovered with a similar median bias and spread, even the projected enclosed DM mass. On the other hand, when considering the projected quantities within $R_\text{Ein}$, Model4 shows better recovery. The projected enclosed DM mass in Model4 presents a $16\%$ better recovery in the median error compared to Model1, while the projected DM fraction shows an improvement of $\sim 10\%$.

When evaluating the total density slope, we found only slight differences between Model1 and Model4 using Dyn and dyLens approaches. It seems that, regardless of the mass model restrictions or the number of kinematical tracers used, the total density slope is consistently well recovered. This is not entirely surprising, as both models also accurately recover the intrinsic total enclosed mass. This is valid for all four measurements that we consider for the total density slope.

Despite the improvement in the enclosed DM quantities reached by the more rigid assumption about the DM scale radius, the DM inner slope remains poorly constrained by both dynamical-only and joint models. The median fractional difference for the Dyn and dyLens Model4 is consistent with that presented in Figure \ref{fig:dm_parameters_recover}. This shows that the rigid assumption about the DM scale radius was not able to reduce the systematics in the recovery of the DM inner slope, although it does improve the recovery of the integrated DM quantities.

\section{Summary and Conclusion}\label{sec:conclusions}
In this work, we present an overview of the self-consistent modelling of SGL and stellar dynamics through the implemented framework \texttt{dyLens}: dynamical and lens modelling. The \texttt{dyLens} framework solves, for a given mass profile parametrised by the MGE formalism, the lens and the Jeans axisymmetric equations simultaneously, allowing for the determination of the lens mass profile and the source reconstruction, helping to break inherent degeneracies in the individual approaches. 

We have tested the systematic errors in the self-consistent modelling, applying the method to a sample of simulated observations created from the Illustris-TNG50. The sample was selected to contain ETGs and to resemble observations carried out by HST and MUSE. Our goal is to replicate the intricate features observed in both lensing and IFU observations. We also compared the results of the self-consistent modelling with results obtained from dynamic and lens-only modelling. We find that:

\begin{itemize}
    \item The combined dyLens model consistently outperforms the other models, exhibiting smaller median fractional errors. Notably, it recovers the intrinsic total mass within $2.5R_\text{eff}$ with a $2\%$ accuracy. 

    \item All models (combined, dynamical and lens) underestimate the DM inner slope, with a median fractional error of $\sim -30\%$. And even though the dyLens model shows a median accuracy of $-1\%$ for the recovered scale radius, the spread is significantly large, making it difficult to draw strong conclusions. 
    
    \item The stellar mass-to-light ratio is well recovered by all models. In particular, the dynamical-only model achieves an accuracy of $1\%$, while the dyLens model achieves $11\%$.

    \item The Einstein radius is robustly recovered for both the lens-only and dyLens models. The lens-only model recovered it with an accuracy of $-5\%$, while the dyLens presents a median fractional difference consistent with zero. 

    \item The total mass density slope is well constrained by both the dynamical-only and dyLens model, regarding the estimation being used, i.e., the average slope or the power-law fit. The accuracy of the recovery is typically of the order of $5\%$ with an intrinsic spread of $\sim 10\%$.

    \item The number of kinematical constraints tested ($15, 35, 55$ bins) does not impact the model outcomes, with all quantities being recovered with similar accuracy and spread.

    \item Typically, integrated quantities involving DM, and the DM parameters themselves, are poorly recovered by all models. However, imposing more restrictive assumptions on the DM halo, such as fixing the scale radius, could alleviate some of the issues. For instance, the projected enclosed DM mass presents an improvement of  $16\%$ in the median fractional error when comparing Model1 (scale radius free) with Model4 (scale radius fixed). However, it is important to note that the spread around these values is still large.
\end{itemize}

Additionally, we employed the pipeline detailed in section \ref{sec:theory} to model the entire sample, which turns the modelling process automatic and uniform. This enables the modelling of larger samples, akin to those anticipated in current/forthcoming surveys such as Euclid, LSST, and CSST. Moreover, given the flexibility of the \texttt{dyLens} framework, this pipeline can be tailored to provide more accurate descriptions of the lens systems under investigation, potentially yielding even more promising outcomes.

\appendix
\section{Sample properties}\label{ap:sample}
Figure \ref{fig:TNG50_sample} presents the final mock sample images and kinematical maps. Note that in the central region of the lens, where the number of kinematical bins is greater, the galaxies are velocity-dispersion-dominated. Table~\ref{tab:sample_features} show some of the main characteristics of the sample. The columns, from left to right, display the subhalo's unique ID at snapshot 84, the inclination angle, the rotational angle along the shortest axis, the total stellar mass, the total DM fraction, the mass-to-light ratio, the triaxiality parameter of the stars and the stellar anisotropy parameter. The DM fraction assumes that the overall mass of the subhalo results from the sum of both stellar and DM masses. 

The triaxiality parameter \citep[][]{Franx1991,Binney2008} is given by

\begin{equation}
    T_\text{star} = \frac{a^2 - b^2}{a^2 - c^2} = \frac{1 - q^2}{1 - s^2},
\end{equation}
where $a$, $b$, and $c$ represent the ellipsoid axis. An ellipsoid is considered perfectly oblate when $T_\text{star} = 0$, while $T_\text{star} = 1$ means a perfectly prolate ellipsoid. While there are slightly differing definitions in the literature for exact intervals, a widely accepted categorisation is as follows: oblate if $0 \leq T_\text{star} < 0.3$; triaxial if $0.3 \leq T_\text{star} < 0.7$; and prolate if $0.7 \leq T_\text{star} \leq 1$. Note that many of the subhalos are not classified as oblate, breaking the assumption made in order to obtain equation \ref{eq:MGE mass density}. But despite that, the models seem to reproduce many of the mass quantities, showing the robustness of the modelling even when some of the assumptions are not strictly followed by the data. 

\begin{table*}
\centering
\begin{filecontents*}{Figs/sample_infos.txt}
subhaloID,$i(^\circ)$,$\phi(^\circ)$,$M_\text{star}$ (M$_\odot$),$f_\text{DM}$,$\Upsilon_\star$ (M$_\odot$/L$_\odot$),$T_\text{star}$,$\beta_z$,$R_\text{eff}$($''$)
7,83.33,55.65,7.34$\times10^{10}$,0.80,1.73,0.46,0.44,0.97
8,84.98,45.31,6.06$\times10^{10}$,0.85,2.49,0.68,0.40,0.54
9,72.09,17.95,6.58$\times10^{10}$,0.77,2.21,0.15,0.40,0.94
11,71.04,20.19,4.71$\times10^{10}$,0.85,2.46,0.70,0.43,0.43
20,67.12,142.40,2.40$\times10^{10}$,0.76,2.92,0.10,0.21,0.17
56405,67.60,65.55,5.03$\times10^{10}$,0.83,2.67,0.39,0.30,0.36
56406,73.31,19.51,2.78$\times10^{10}$,0.64,2.71,0.81,0.52,0.31
83991,79.08,0.09,5.31$\times10^{10}$,0.84,2.03,0.67,0.47,0.51
83996,76.49,65.47,3.39$\times10^{10}$,0.83,1.99,0.85,0.47,0.34
84010,83.13,126.63,1.93$\times10^{10}$,0.69,2.15,0.85,0.54,0.24
100675,69.36,162.06,3.34$\times10^{10}$,0.52,2.23,0.22,0.32,0.39
116278,71.62,87.33,7.20$\times10^{9}$,0.76,2.24,0.90,0.56,0.16
172209,70.51,62.08,6.59$\times10^{10}$,0.66,1.86,0.35,0.48,0.91
313415,67.73,23.86,1.83$\times10^{11}$,0.97,2.58,0.17,0.23,0.35
341482,79.13,17.41,1.93$\times10^{11}$,0.94,1.94,0.47,0.41,0.98
344595,81.38,30.60,1.24$\times10^{11}$,0.95,2.55,0.37,0.43,0.69
396742,80.86,144.70,1.35$\times10^{11}$,0.94,2.70,0.45,0.38,0.49
414107,65.63,153.69,1.19$\times10^{11}$,0.94,2.26,0.46,0.30,1.10
451938,72.88,77.19,3.50$\times10^{10}$,0.94,2.31,0.67,0.58,0.36
485608,84.73,60.38,4.31$\times10^{10}$,0.96,2.13,0.61,0.51,0.41
545285,83.38,26.87,3.26$\times10^{10}$,0.93,2.20,0.72,0.56,0.36
\end{filecontents*}
\csvreader[no head,tabular=|l|c|c|c|c|c|c|c|,
  table head=\hline,late after line=\\\hline]{Figs/sample_infos.txt}
  {1=\one,2=\two,3=\three,4=\four,5=\five,6=\six,7=\seven,9=\nine}
  {\one & \two & \three & \four & \five & \six & \seven & \nine}%
\caption{TNG50 sample properties. From left to right, columns are the subhalo's unique ID at snapshot 84, the inclination angle, the rotational angle along $z$-axis, the total stellar mass, the total DM fraction, the mass-to-light ratio, the triaxiality parameter of the stars, and the MGE effective radius.}
\label{tab:sample_features}
\end{table*}

\begin{figure*}
\centering
    \includegraphics[scale=0.65]{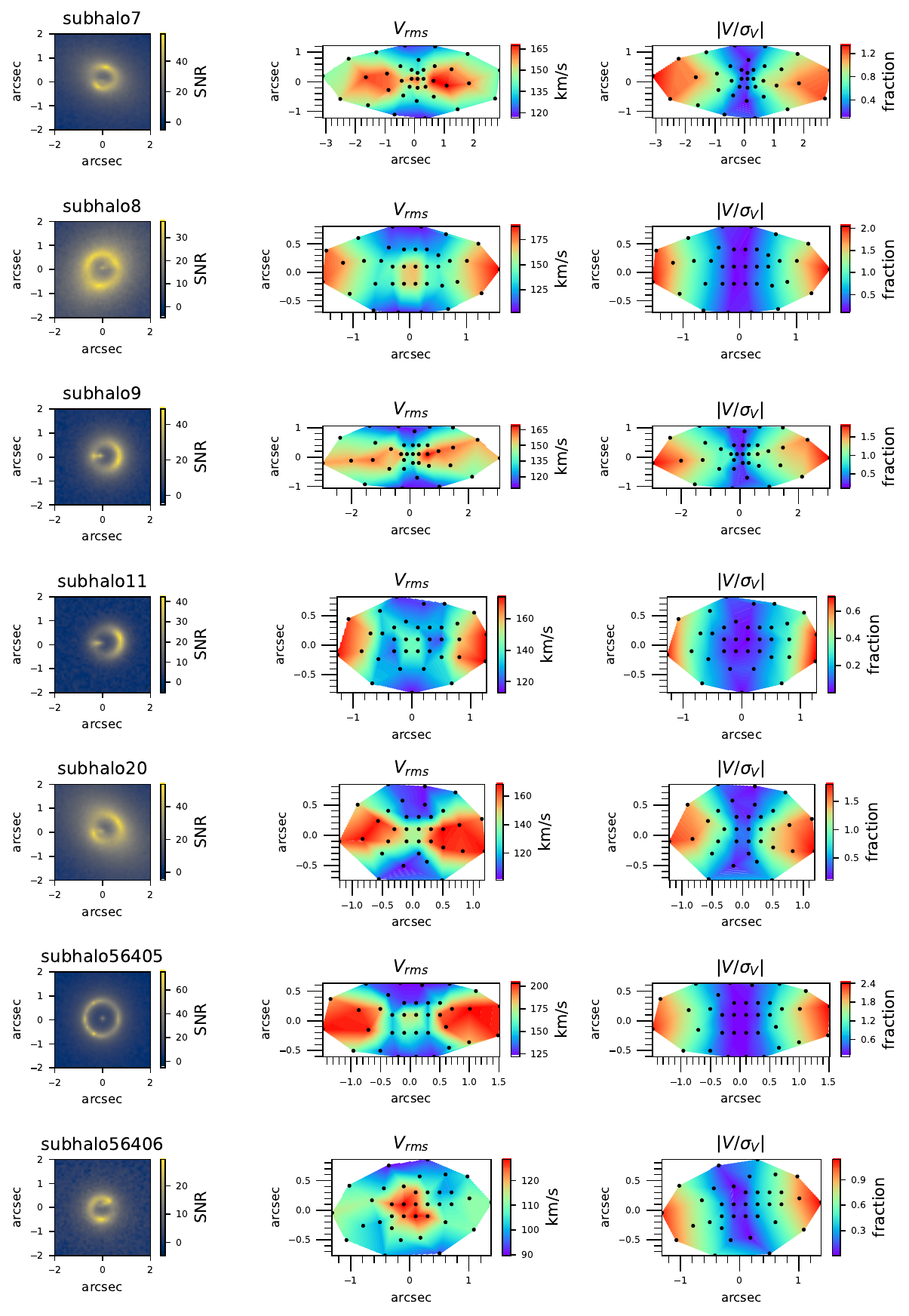}
    \caption{TNG50 mock sample. The leftmost column displays the deflected source, the middle column is the IFU kinematical data ($\sim35$bins), and the rightmost column illustrates the modulus ratio of the velocity by the velocity dispersion. In the lensing data, the colour bar corresponds to the SNR. The black dots on the kinematical maps correspond to the centres of the Voronoi bins.
}
    \label{fig:TNG50_sample}
\end{figure*}

\begin{figure*}
\centering
    \includegraphics[scale=0.65]{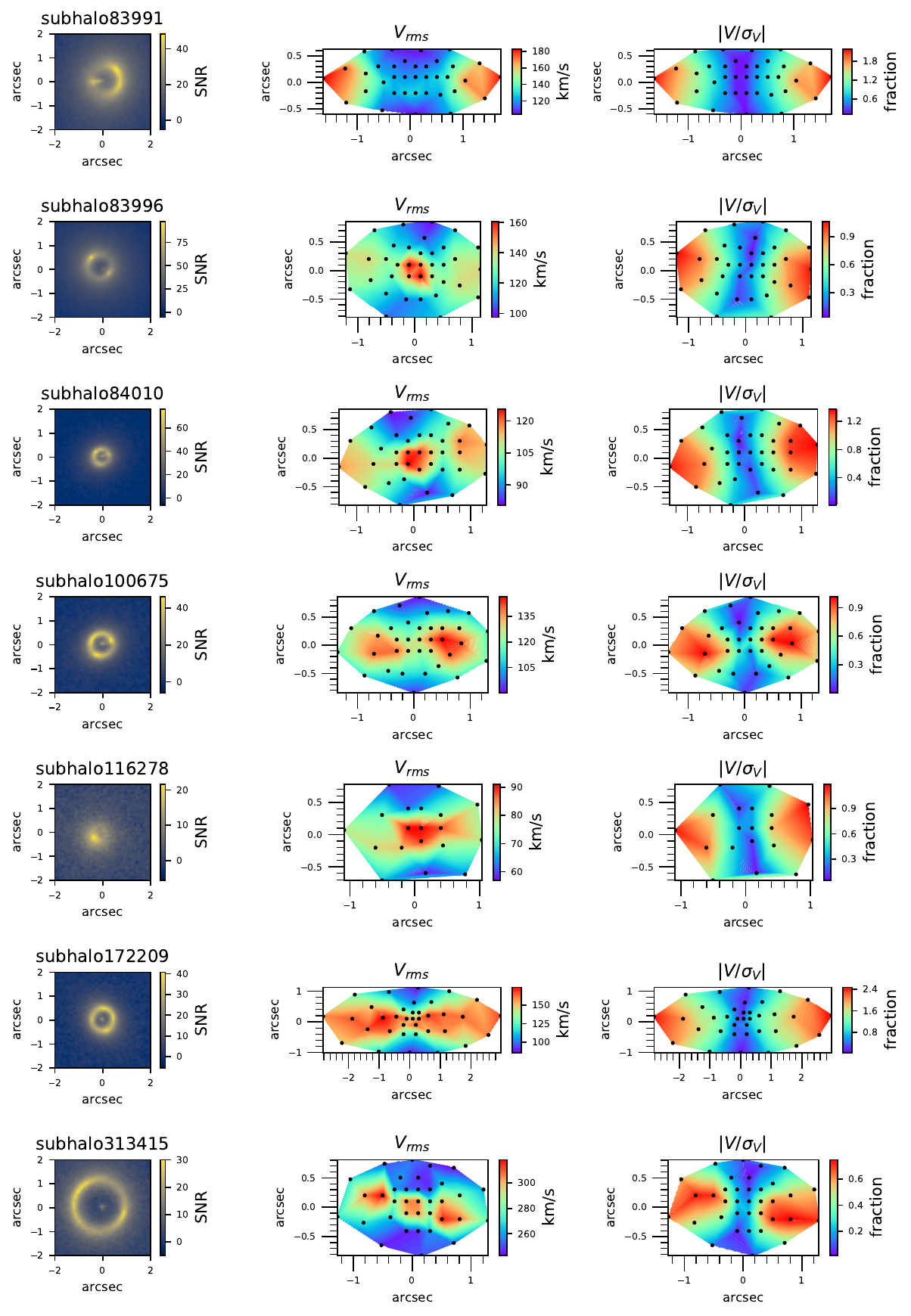}
    \caption{Continuation of Figure \ref{fig:TNG50_sample}.
}
\end{figure*}

\begin{figure*}
\centering
    \includegraphics[scale=0.65]{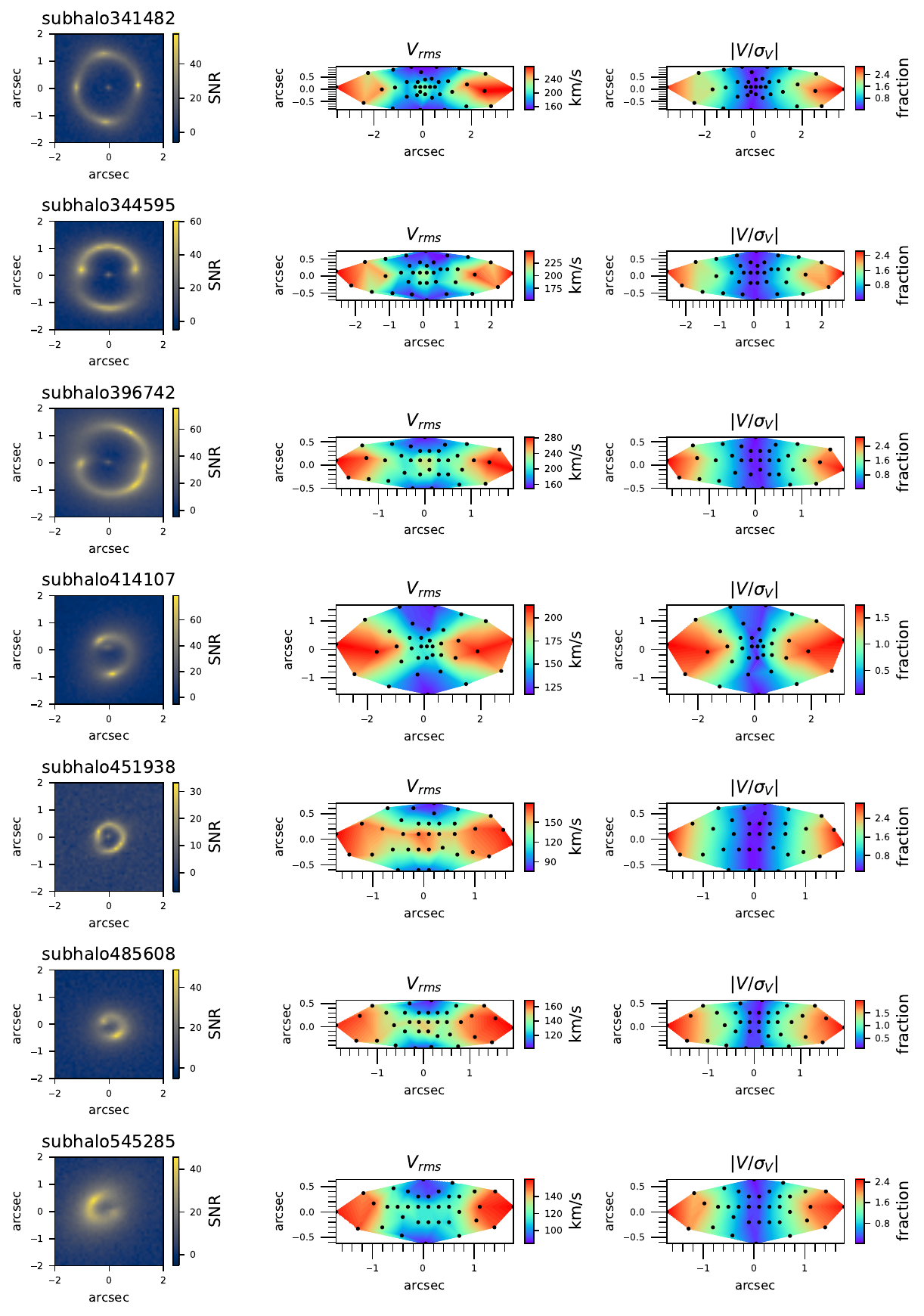}
    \caption{Continuation of Figure \ref{fig:TNG50_sample}.
}
\end{figure*}

\section{Kinematical uncertainties}\label{ap:kin_unc}
To ensure reliable uncertainties in the velocity maps, we create mock galaxy spectra to derive realistic uncertainties in the same way as the uncertainties are derived from real observations. As described in section \ref{sec:sample}, the mock spectra were created such that they resemble MUSE observations, and the fitting procedure carried is similar to that applied to real data. 

Drawing from previous investigations of ETGs by refs. \cite{Harrison2010} and \cite{McDermid2015}, we used three bins of age ([$2.5$, $6.3$, $12.6$] Gyrs) and three bins of metallicity ([Z/H] = [$-0.4$, $0.0$, $0.22$]) to construct mock galaxy spectra. We use the E-MILES SSP models \citep{Vazdekis2016} that best encompass these parameter ranges. Given our primary interest in ETGs, we explore four distinct velocity dispersion values, namely $\sigma_v$ = [$250$, $280$, $300$, $350$]\,km/s, while maintaining a fixed velocity of $40$\,km/s. This choice ensures that all simulated spectra are velocity dispersion-dominated, which aligns with the expected behaviour for ETGs. Furthermore, we examine four SNR levels for the spectrum ([$15$, $20$, $25$, $35$]), aiming to assess whether lower SNR might introduce biases in the recovery of kinematical information. 

We then fit all the $144$ mock spectra, encompassing a variety of ages, metallicities, velocity dispersions and SNR values with the \texttt{pPXF} algorithm using the Indo-US templates \citep[][]{Indo-US}. Neither additive nor multiplicative polynomials were incorporated into the fitting process since they are usually used to correct mismatches between the fit and the data, and here we want to avoid introducing too many sources of variation. 

\begin{figure}[h!]
\centering
	\includegraphics[scale=0.45]{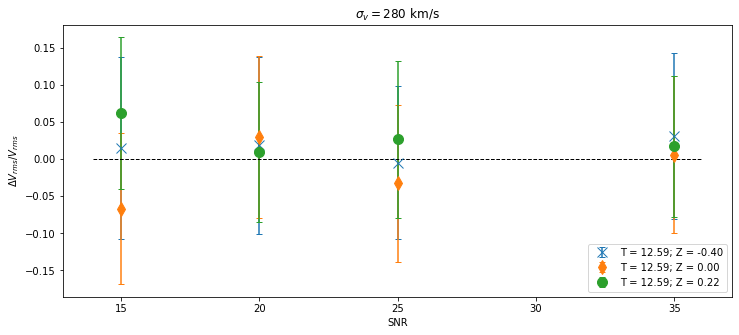}
    \includegraphics[scale=0.45]{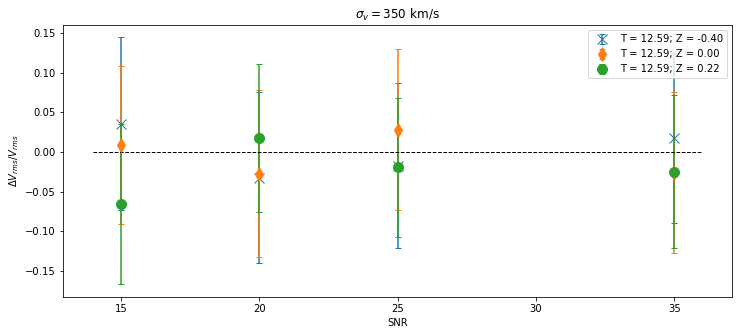}
    \caption{The fractional uncertainty on $v_{\text{rms}}$ as a function of the SNR for a fixed age of $12.59$\,Gyrs and a fixed input velocity dispersion of $\sigma_v = 280$\,km/s (top) and  $\sigma_v = 350$\,km/s (bottom). Different colours represent different metallicities as indicated in the inset plot. 
}
    \label{fig:ppxf-recover1}
\end{figure}

No evident bias was observed in any configuration, and all combinations successfully retrieve the input $v_\text{rms}$ within the associated uncertainties. This is illustrated by Figure~\ref{fig:ppxf-recover1}, where the fractional uncertainty ($\frac{\text{model - data}}{\text{data}}$) as a function of the SNR, for a fixed age and varying metallicity are shown. All inputs are well recovered, regarding SNR, metallicity or input $\sigma_v$. Similar plots show that this behaviour is consistent for different ages and $\sigma_v$ as well. 

\begin{figure}[h!]
\centering
    \includegraphics[scale=0.35]{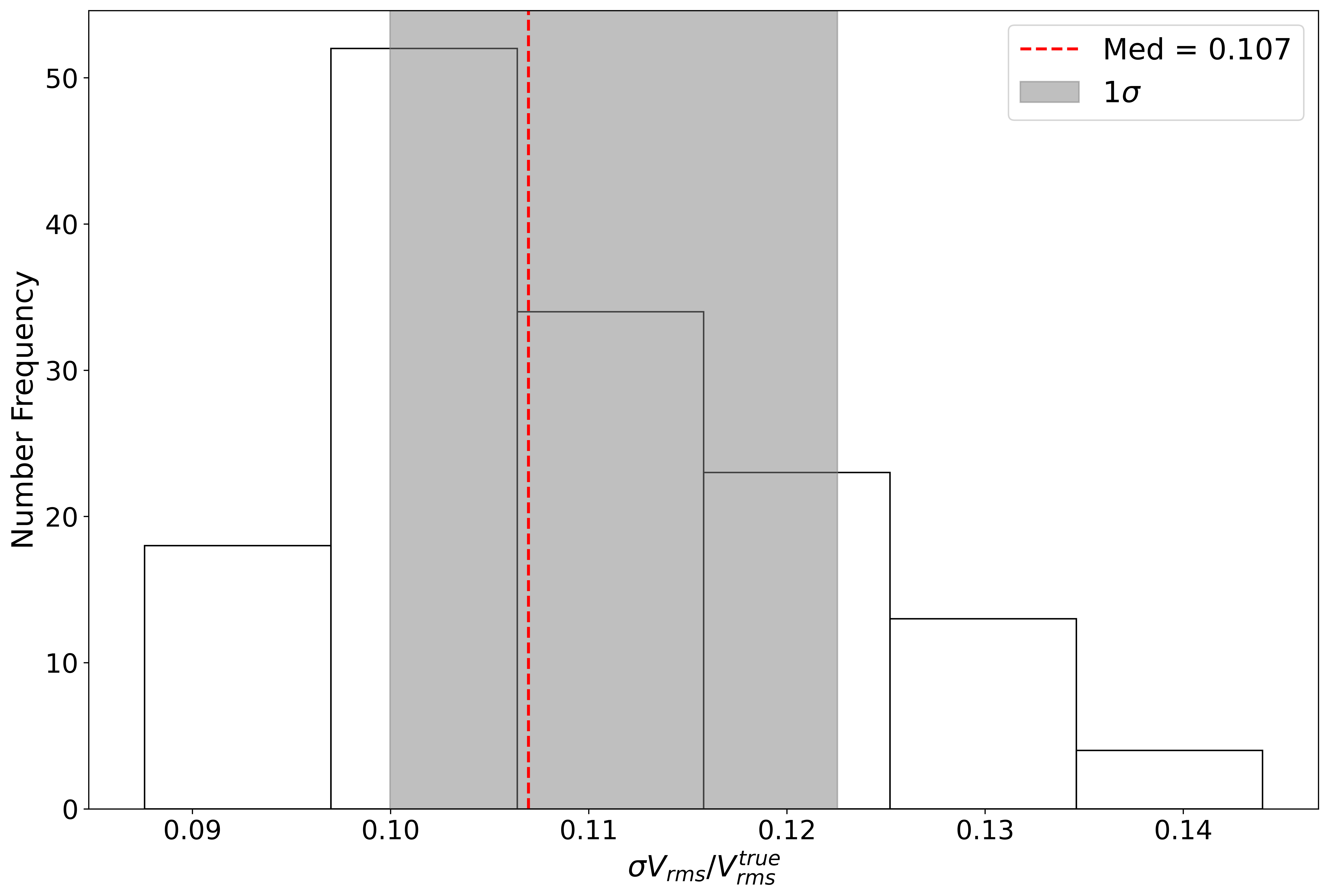}
    \caption{The ratio of the uncertainty on $v_\text{rms}$ by the true  $v_\text{rms}$ value. The median value of $11\%$ is the value assumed to derive the uncertainties in the mock IFU data. This distribution takes all the 144 simulated spectra into account.
}
    \label{fig:vrms_dist_u}
\end{figure}

In Figure~\ref{fig:vrms_dist_u}, we show the ratio of the uncertainty in $v_\text{rms}$ to the true $v_\text{rms}$ value for the $144$ simulated spectra. The dashed line marks the median value ($11\%$) and the red shaded area represents $1\sigma$ credible interval.

\section{Mass model parameters and Priors}\label{ap:mass_model}

Table \ref{table:TNG50_priors} describes the parameters and the priors applied to model the simulated sample. For the Lens and dyLens models, the priors are updated according to the pipeline described in section \ref{sec:theory}.

\begin{table*}
\centering
 \caption{Mass model parameters and priors applied to the simulated galaxy sample. From left to right, the columns are the parameter symbol, prior applied during Ph1/Dyn, parameter description and physical unit. $^{(*)}$ minimum inclination allowed by equation (\ref{eq:_q_deproj}).}
\label{table:TNG50_priors}
 \begin{tabular}{lccc}
  \hline
 Parameter  & Prior & Description & Physical Unit\\
\hline
$x^\prime_0$           & $\mathcal{N}[0.0, 0.3]$   &   \makecell{Source $x^\prime$ centre} & arcsec\\  &&& \\
$y^\prime_0$           & $\mathcal{N}[0.0, 0.3]$   &   \makecell{Source $y^\prime$ centre} & arcsec\\  &&& \\
$q_{\text{source}}$ & $\mathbf{U}[0.1, 1.0]$   &   \makecell{Source axial ratio} & - \\  &&& \\
$\phi_{\text{source}}$ & $\mathbf{U}[0.0, 180.0]$   &   \makecell{Source orientation  \\ angle counterclockwise \\ from $x^{\prime}$-axis} & degree \\  &&& \\
$I_{\text{source}}$ & $\log_{10}\mathbf{U}[10^{-6}, 10^{6}]$   &   \makecell{Source intensity} & counts/s \\  &&& \\
$R_\text{eff}$& $\mathbf{U}[0.0, 30.0]$   &   \makecell{Effective radius} & arcsec \\  &&& \\
$n$& $\mathbf{U}[0.5, 8.0]$   &   \makecell{Sérsic index} & -\\  &&& \\
$\Upsilon$           & $\mathbf{U}[1.0, 10.0]$   &   \makecell{Constant $M/L$} & M$_\odot$/L$_\odot$\\  &&& \\
 $\beta_z$            & $\mathbf{U}[-0.5, 0.5]$  &   Anisotropy & -\\  &&& \\
 $i$       & $\mathbf{U}[\text{min.}^{(*)}, 90.0]$   &   \makecell{Galaxy inclination}  & degree\\  &&& \\
 $\log_{10}{\frac{\rho_s}{\text{M}_\odot\,\text{pc}^{-3}}}$             & $\mathbf{U}[-6.0, 0.0]$   &   \makecell{$\log_{10}$ of the\\ characteristic\\density at $r_s$}  & -\\  &&& \\
 $r_s$                  & $\mathbf{U}[0.01, 30.0]$  &   \makecell{Scale radius \\ of DM halo} & arcsec\\  &&& \\
  $\gamma_\text{DM}$                  & $\mathbf{U}[0.5, 2.0]$  &   \makecell{Inner slope \\ of the DM halo} & -\\  &&& \\
 $\text{shear}_{\text{mag}}$ & $\mathbf{U}[0.0, 1.0]$ &    \makecell{Shear magnitude} & -\\ &&& \\
 $\text{shear}_{\phi}$  & $\mathbf{U}[0.0, 180.0]$    &    \makecell{Shear angle \\ counterclockwise \\ from $x^\prime-$axis}& degree\\ &&& \\
 \hline
 \end{tabular}
\end{table*}

\acknowledgments
The authors thank the referee for their comments and suggestions, which led to an improved version of the manuscript. This project is funded by Conselho Nacional de Desenvolvimento Científico e Tecnológico (CNPq). 
CF ackwoledges funding from CNPq  and the Rio Grande do Sul Research Foundation (FAPERGS) through grants CNPq-315421/2023-1 and FAPERGS-21/2551-0002025-3.
ACS acknowledges funding from CNPq and FAPERGS through grants  CNPq-314301/2021-6 and FAPERGS/CAPES 19/2551-0000696-9. 
The authors acknowledge the National Laboratory for Scientific Computing (LNCC/MCTI, Brazil) for providing HPC resources of the SDumont supercomputer, which have contributed to the research results reported within this paper. URL: \url{http://sdumont.lncc.br}. This  work made use of the CHE cluster, managed and funded by COSMO/CBPF/MCTI, with financial  support  from  FINEP  and  FAPERJ,  and  operating  at  the  Javier  Magnin  Computing Center/CBPF. The IllustrisTNG simulations were undertaken with compute time awarded by the Gauss Centre for Supercomputing (GCS) under GCS Large-Scale Projects GCS-ILLU and GCS-DWAR on the GCS share of the supercomputer Hazel Hen at the High Performance Computing Center Stuttgart (HLRS), as well as on the machines of the Max Planck Computing and Data Facility (MPCDF) in Garching, Germany. CMC thank Wolfgang Enzi and Lindsay Oldham for the fruitful discussion. The authors thank Ling Zhu for the discussion about dynamical modelling.

\paragraph{Data Availability.} The IllustrisTNG simulations are publicly available at \url{https://www.tng-project.org/data/}. Additional data directly related to this publication will be available at \url{https://github.com/carlosRmelo} or on request from the corresponding author.

\paragraph{Software Citation.}

\texttt{Astropy} \citep{astropy}; \texttt{dynesty} \citep{Speagle2020,sergey_koposov_2023_7995596}; \texttt{Jampy} \citep{Cappellari2008,Cappellari2020}; \texttt{Jupyter} \citep{jupyter}; \texttt{Matplotlib} \citep{Hunter:2007}; \texttt{MgeFit}  \citep{Cappellari2002}; \texttt{Numba} \citep{numba}; \texttt{NumPy}   \citep{numpy}; \texttt{pPXF} \citep{Cappellari2004}; \texttt{PyAutoLens} \citep{Nightingale2018,Nightingale2021}; \texttt{SciPy}   \citep{Scipy_2020}; \texttt{VorBin} \citep{Cappellari2003}; \texttt{lenstronomy} \citep{Birrer2018, Birrer2021}; \texttt{illustris-tools} \citep{Li2016}; \texttt{Sim\_MaNGA\_lens}\footnote{\url{https://github.com/caoxiaoyue/Sim_MaNGA_lens}} \cite{Cao2022}




\bibliography{main.bib}{}
\bibliographystyle{JHEP}





\end{document}